\documentclass[12pt]{emulateapj}

\def\vv{\hbox{\it V\/}}

\def\umv{\hbox{\it U--V\/}}
\def\bmi{\hbox{\it B--I\/}}

\def\vmi{\hbox{\it V--I\/}}

\def\Min{${}^{\prime}$\llap{.}}
\def\Sec{${}^{\prime\prime}$\llap{.}}
\def\deg{${}^\circ$}

\newcommand{\kms}{\mbox{km s$^{-1}$}}

\def\Sec{${}^{\prime\prime}$\llap{.}}

\def\Min{${}^{\prime}$\llap{.}}

\shorttitle{The Carina Project. IV. radial velocity distribution}
\shortauthors{Fabrizio et al.}

\begin{document}
\title{The Carina Project. IV. radial velocity distribution\altaffilmark{1}}

\author{
M.~Fabrizio\altaffilmark{2},
M.~Nonino\altaffilmark{3},
G.~Bono\altaffilmark{2,4,5},
I.~Ferraro\altaffilmark{4},
P.~Fran\c{c}ois\altaffilmark{6},
G.~Iannicola\altaffilmark{4},
M.~Monelli\altaffilmark{7},
F.~Th\'evenin\altaffilmark{8}
P.~B.\ Stetson\altaffilmark{9,14},
A.~R.\ Walker\altaffilmark{10},
R.~Buonanno\altaffilmark{2,11},
F.~Caputo\altaffilmark{4},
C.~E.\ Corsi\altaffilmark{4},
M.~Dall'Ora\altaffilmark{12},
R.~Gilmozzi\altaffilmark{5},
C.R.~James\altaffilmark{13},
T.~Merle\altaffilmark{8},
L.~Pulone\altaffilmark{4},
M.~Romaniello\altaffilmark{5}
}

\altaffiltext{1}{Based on spectra collected with the spectrograph FORS2 available at the ESO
Very Large Telescope (VLT), Cerro Paranal, (072.D-0671(A) PI:~Bono -- 078.B-0567(A) PI:~Th\'evenin)}

\altaffiltext{2}{Dipartimento di Fisica, Universit\`a di Roma Tor Vergata, via della Ricerca
Scientifica 1, 00133 Rome, Italy; michele.fabrizio@roma2.infn.it}
\altaffiltext{3}{INAF--Osservatorio Astronomico di Trieste, via G.B. Tiepolo 11, 40131 Trieste, Italy}
\altaffiltext{4}{INAF--Osservatorio Astronomico di Roma, via Frascati 33, Monte Porzio Catone, Rome, Italy}
\altaffiltext{5}{European Southern Observatory, Karl-Schwarzschild-Str. 2, 85748 Garching bei Munchen, Germany}
\altaffiltext{6}{Observatoire de Paris-Meudon, GEPI, 61 avenue de l'Observatoire, 75014 Paris, France}
\altaffiltext{7}{Instituto de Astrof\'isica de Canarias, Calle Via Lactea, E38200 La Laguna, Tenerife, Spain}
\altaffiltext{8}{Observatoire de la C\^ote d'Azur, BP 4229, 06304 Nice, France}
\altaffiltext{9}{Dominion Astrophysical Observatory, Herzberg Institute of Astrophysics, National Research Council, 5071 West Saanich Road, Victoria, BC V9E 2E7, Canada}
\altaffiltext{10}{Cerro Tololo Inter-American Observatory, National Optical Astronomy Observatory, Casilla 603, La Serena, Chile}
\altaffiltext{11}{Agenzia Spaziale Italiana--Science Data Center, ASDC c/o ESRIN, via G. Galilei, 00044 Frascati, Italy}
\altaffiltext{12}{INAF--Osservatorio Astronomico di Capodimonte, via Moiariello 16, 80131 Napoli, Italy}
\altaffiltext{13}{Department of Physics, Sam Houston State University, Huntsville, Texas 77341, USA}
\altaffiltext{14}{Visiting Astronomer, Cerro Tololo Inter-American Observatory,
National Optical Astronomy Observatories, operated by AURA, Inc., under cooperative agreement with the NSF.}

\date{\centering drafted \today\ / Received / Accepted }

\begin{abstract}
We present new and accurate radial velocity (RV) measurements of luminous stars
of all ages (old horizontal branch, intermediate--age red clump, and young blue
plume, as well as red giants of a range of ages; 20.6$\leq$\vv$\leq$22) in the
Carina dwarf spheroidal galaxy, based on low-resolution spectra collected with
the FORS2 multi-object slit spectrograph at the VLT. This data set was
complemented by RV measurements based on medium and high-resolution archive
spectra of brighter (\vv$\lesssim$20.6) Carina targets collected with the
GIRAFFE multi-object fiber spectrograph at the VLT. 
The combined sample includes more than 21,340 individual spectra of
$\approx$2,000 stars covering the entire body of the galaxy. The mean
($<$RV$>$=220.4$\pm$0.1 \kms) and the dispersion ($\sigma$=11.7$\pm$0.1 \kms) of
the RV distribution of candidate Carina stars ($\sim$1,210 objects,
180$\le$$RV$$\le$260 \kms, 4$\sigma$) agree quite well with similar measurements
available in the literature. To further improve the statistics, the accurate RV
measurements recently provided by \citet{walker07} were also added to the
current data set. We ended up with a sample of $\sim$1,370 RV measurements of
candidate Carina stars that is $\approx$75\% larger than any previous Carina RV
sample.
We found that the hypothesis that the Carina RV distribution is Gaussian can be
discarded at 99\% confidence level. The mean RV across the body of the galaxy
varies from $\sim$220 \kms\ at a distance of 7\arcmin\ ($\sim$200 pc) from the
center to $\sim$223 \kms\ at 13\arcmin\, ($\sim$400 pc, 6$\sigma$ level) and
flattens out to a constant value of $\sim$221 \kms\ at larger distances
(600 pc, 4$\sigma$ level).
Moreover and even more importantly, we found that in the Carina regions where
the mean RV is smaller the dispersion is also smaller, and the RV distribution
is more centrally peaked (i.e. the kurtosis attains larger values). The
difference in mean RV is more than 4 \kms\ (9$\sigma$ level), when moving from E to
W, and more than 3 \kms\ ($\sim$7$\sigma$ level), when moving from N to S. 
The RV gradient appears to be in the direction of the Carina proper motion.
However, this parameter is affected by large uncertainties to reach a firm
conclusion. There is evidence of a secondary maximum in RV across the Carina
center ($|D|$$\le$200 pc). The occurrence of a secondary feature across the
Carina center is also supported by the flat-topped radial distribution based on
the photometric catalog. These findings are reminiscent of a substructure with
transition properties, already detected in dwarf ellipticals, and call for
confirmation by independent investigations.
\end{abstract}


\section{Introduction}
Dwarf galaxies are fundamental laboratories to investigate the influence of the
environment on star formation and on chemical evolution in stellar systems that
are several order of magnitudes smaller than giant galaxies. Empirical evidence
indicates that in the Local Group together with the dwarf ellipticals (dE,
M32-like) and the dwarf spirals (dS, M33-like) we are facing three different
dwarf morphological types. The dwarf spheroidals (dSphs) show either single
(Cetus, \citealt{monelli10a}; Tucana, \citealt{monelli10b}) or multiple star
formation events (Carina, \citealt{bono10}), are spheroidal in morphology and
devoid of neutral hydrogen \citep{bouchard05}. The dwarf irregulars (dIs) host a
mix of old and young stellar populations, but they have recently experienced an
intense SF episode \citep{sanna09,cole09}, moreover, they have irregular
morphology and host a significant fraction of neutral hydrogen. The transition
dwarfs have properties between the dSphs and the dIs. The observational scenario
concerning the dwarfs in the LG was recently enriched by the discovery of
ultra-faint dSphs in the SDSS \citep{zucker06,tolstoy09,wyse10}, but their
properties still need to be investigated in detail. Although the morphological
classification appears robust, we still lack firm quantitative constraints
concerning the evolution/transition between different dwarf morphological types.
\citet{kormendy85,kormendy87} found a well defined dichotomy between ellipticals
(Es) and spheroidals (Sphs, \citealt{kormendy09}). Early type galaxies are
distributed along a sequence moving from cD to dEs, while spheroidals are
distributed along a sequence that overlaps with spirals and dwarf irregulars
(see Fig.~1 in \citealt{kormendy09}). This empirical dichotomy seems to suggest
that E and Sph galaxies are stellar systems that underwent different formation
and evolution processes. In particular, the Sphs might be either spirals or
irregulars that lost their gas or transformed it into stars. However, there is
no general consensus concerning the E--Sph dichotomy, since the correlation
between the shape of the brightness profile and the galaxy luminosity is
continuous when moving from Es to Sphs \citep{jerjen97,gavazzi05,ferrarese06}.

Theoretical and empirical uncertainties affect our understanding of several
empirical correlations. More than twenty years ago it was suggested by
\citet{skillman89} that the luminosity of dwarfs is correlated with their mean
metallicity. However, it is not clear whether this correlation is linear over
the entire metallicity range or shows a plateau in the metal-poor regime
\citep{helmi06,mateo08,kirby08}. More recently, it was suggested by
\citet{woo08} that dEs/dSphs in the LG, at fixed total visual luminosity, show
larger mass-to-light (M/L) ratios (see their Fig.~1). The same systems are, at
fixed total stellar mass, more metal-rich than dIs (see their Fig.~10 and
\citealt{mateo98araa}).

Precise and homogeneous photometric and spectroscopic data are required to
address the above open issues concerning the formation and evolution of these
systems. The advent of multi-object spectrographs at the 8-10m class telescopes
provided the unique opportunity to collect sizable samples of low-, medium- and
high-resolution spectra covering the entire body of nearby dwarf galaxies. By
using hundreds of radial velocity (RV) measurements it has been suggested by
\citet{kleyna04} that Sextans dSph host kinematic substructures, in particular,
they found that the RV dispersion of the stellar population located across the
center of this system is close to zero and increases outside the core. A
different kinematic status has also been suggested by \citet{battaglia08} for
the two distinct stellar populations in the Sculptor dSph. 

Moreover, evidence of a fall-off in the velocity dispersion at large galactic
distances was suggested in Sextans, Draco and Ursa Minor
\citep{kleyna03,wilkinson04,wilkinson06}. However, \citet{walker09}, using 
large and homogeneous samples of RV measurements based on high-resolution 
spectra collected with Mike Multi Fiber Spectrograph (MMFS) at Magellan 
\citep[hereinafter W07]{walker07}, found that the RV profile in several dSphs 
(Carina, Draco, LeoI, LeoII, Ursa Minor) are flat at large projected radii 
(R$\ge$1 kpc), while Fornax, Sculptor and Sextans show a gentle decline. 

The observational scenario is far from being settled, and indeed,
\citet{lokas08} and \citet{lokas09} using the same RV measurements provided by
\citet{walker07}, but a different algorithm to reject tidally stripped stars,
found a significant decline in the RV dispersion profile of Carina, Fornax,
LeoI, Sculptor and Sextans. Furthermore, it was also suggested that kinematic
status of Carina \citep{munoz06} and Bootes \citep{belokurov06} might be
disturbed by the Milky Way. In this context, Carina plays a key role because it
is relatively close ($\mu$=20.10 mag), shows at least two well separated SF
episodes (t=1-6 and 12 Gyr) and it is metal-poor ($<$[Fe/H]$>\sim$--1.7 dex).

Our group is involved in a long-term project on the evolutionary properties,
variable stars, kinematic and abundances of stellar populations in the Carina
dSph. This is the fourth paper of a series and it is focused on the radial
velocity distribution of candidate Carina stars. The structure of the paper is
as follows. In Section 2, we discuss in detail the different spectroscopic data
sets we collected for this experiment, together with the approach we adopted to
reduce the data. Section 3 deals with radial velocity measurements and with the
observational strategy we adopted to validate radial velocities based on low-,
medium- and high resolution spectra. In Section 4 we lay out the selection
criteria adopted to identify candidate Carina stars and their radial velocity
distribution. In Section 5, we investigate the velocity distribution, the
velocity dispersion and the kurtosis of candidate Carina stars as a function of
the projected radial distance. In this section, we also address the similarity
between spectroscopic and photometric radial distributions. Finally, in Section
6 we summarize the results of this investigation and briefly outline possible
future extensions of this photometric and spectroscopic experiment.

\section{Spectroscopic data sets and data reduction}
To collect low-resolution spectra of old (Horizontal Branch [HB], \vv$\sim$20.75
mag), intermediate (Red Clump [RC], \vv$\sim$20.5 mag) and young (Blue Plume
[BP], \vv$\sim$22 mag) stellar tracers (see Fig.~1) we observed five fields with
FORS2 at VLT \citep{appe98} in service mode\footnote{ESO programs 072.D-0671(A),
P.I.:~Bono; 078.B-0567(A), P.I.:~Th\'evenin}. The program was almost completed
by one of us (M.N.) during a Visitor mode observations in December 2004. A
significant fraction of FORS2 spectra were collected using the 1400V grism,
which covers H$_\beta$ ($\lambda$=4861.34 \AA) and MgI Triplet
(MgIT\footnote{MgIT: $\lambda$=5167.32, 5172.68, 5183.62 \AA}) lines, with a
nominal resolution of 2100 ($\lambda_c$=5200 \AA). The slits used were 0\Sec6
length, and typically 10\Sec ~long, with a wavelength coverage
$\approx$4560--5860 \AA. The exposure times range from 22 minutes ({\em short})
to more than one hour ({\em long}). 

During the December 2004 run, the 1028z grism was also used with a wavelength
coverage $\approx$7730--9480 \AA. Its nominal resolution is 2560
($\lambda_c$=8600 \AA) and covers the near-infrared (NIR) calcium triplet
(CaT\footnote{CaT: $\lambda$= 8498.02, 8542.09, 8662.14 \AA}). Using the 1400V
grism we secured in 19.12 hrs of exposure time 743 individual spectra of 
356 stars with DIMM seeing ranging from 0\Sec35 ~to 1\Sec27, while using the
1028z grism we secured in 1 hr of exposure time 113 spectra of 73 stars with
seeing 0\Sec38--0\Sec59. Most of the stars with 1028z spectra also have 1400V
spectra. We ended up with a total sample of 359 stars either with single or
multiple FORS2 spectra. 
%
In particular, the LR data set collected with the grism 1400V includes 356 stars
and among them eleven have a single spectrum, while the others a number of
spectra ranging from two to twelve. The LR data set collected with the grism
1028z includes 73 stars and among them 39 have single spectra, while the others
a number of spectra ranging from two to three. In total the LR data set includes
359 stars of which eleven have a single spectrum and the others a number of
spectra ranging from two to 14. Table~1 gives the log of the observations of the
spectra collected with FORS2. From left to right are listed the date, the
pointing, the coordinates, the grism, the exposure time and the seeing.

These data were complemented by archival medium and high-resolution spectra of
Red Giant (RG) and RC stars collected with GIRAFFE \citep{pas02} at
VLT\footnote{ESO programs 074.B-0415(A), P.I.: Shetrone; 171.B-0520(B), P.I.:
Gilmore; 180.B-0806(B), P.I.: Gilmore} and covering the entire body of the
galaxy (see Fig.~1). The GIRAFFE spectra \citep{koch06} were collected using the
grating LR8 with a nominal resolution of 6,500 centered on the NIR CaT. This
data set (hereinafter GMR03) includes 10,394 individual fiber spectra of 1,070
stars and were secured with a total exposure time of 95.4 hrs with seeing
condition of 0\Sec4--2\Sec2. 
%
Note that this data set was also adopted by \citet{munoz06} in their analysis of
Carina radial velocity distribution.

We also use the high-resolution spectra collected with the gratings: HR10
(R=19,800, 5339$\lesssim$$\lambda$$\lesssim$5619 \AA), HR13 (R=22,500,
6120$\lesssim$$\lambda$$\lesssim$6405 \AA) and HR14A (R=17,740,
6308$\lesssim$$\lambda$$\lesssim$6701 \AA). This data set (hereinafter GHR)
includes 2,002 individual spectra of 98 stars, secured with a total exposure
time of 24 hrs and seeing 0\Sec57--3\Sec20. The above data were complemented by
a new set of GIRAFFE spectra collected with LR8 and including 8,092 individual
spectra of 959 stars. This data set (hereinafter GMR08) was secured with 48.4
hrs of exposure time and seeing of 0\Sec54--1\Sec94. We ended up with a total
sample of 1,931 stars either with single or multiple GIRAFFE spectra. 

The GMR ($<$GMR03$+$GMR08$>$, weighted mean) data set includes 1,887 stars and
only one star has a single spectrum, while the others have multiple spectra
ranging from two to 35. The HR data set includes 98 stars and all of them have
multiple spectra ranging from seven to nine. Moreover, the GHR and the GMR03
spectra have 38 objects in common, while 125 objects are in common between GHR
plus GMR03 and GMR08 spectra. The LR resolution data set has 145 objects in
common with the GIRAFFE (GHR plus GMR) data set. The objects in common were
adopted to calibrate GMR and LR spectra.

Data plotted in the top panels of Fig.~1 show the spatial coverage of the
different data sets, while the bottom panels show the location of the targets in
the \vv,\bmi\ Color-Magnitude Diagram (CMD, \citealt{bono10}). The numbers in
parenthesis give the number of stars for which we estimated the radial velocity
and the total number of stars for which we collected at least one spectrum. 

Data plotted in the bottom panels of Fig.~1 show several interesting features:
{\em i)} {\em Stellar tracers} -- The FORS2 spectra cover old, intermediate and
young stellar tracers (\vv$\lesssim$22, 359 RVs). {\em ii)} {\em Statistics} --
Although the exposure time of the GIRAFFE spectra is a factor of 4-8 longer than
the FORS2 spectra, their limiting magnitude is \vv$\lesssim$20.6 (fibers vs
slits). This means that this data set does not cover young (BP) and truly old
(HB) tracers. The RGs are, indeed, a mix of old and intermediate--age stellar
tracers. However, the sample of RV measurements based on GIRAFFE spectra (1,985
vs 359) is more than a factor of five larger than FORS2 (FoV $\sim$25 vs $\sim$5
arcmin squared). {\em iii)} {\em Multiplicity} -- A significant fraction of our
targets have multiple spectra. This means that we can provide robust estimates
of intrinsic errors. {\em iv)} -- {\em A significant overlap of low, medium and
high-resolution spectra} -- We can constrain the occurrence of possible
systematic errors in RV measurements using spectra collected with different
instruments and different gratings/grisms.

We reduced our FORS2 data using standard IRAF\footnote{IRAF is distributed by
the National Optical Astronomy Observatory, which is operated by the Association
of Universities for Research in Astronomy, Inc., under cooperative agreement
with the National Science Foundation.} tasks, using day time associated bias
frames, flat field and calibration lamps.
The wavelength calibration was carried out using the task {\tt identify}, 
daytime lamp frames and a set of 8-12 calibration lines. We typically end up
with an accuracy better than $\approx$0.02--0.04 \AA, $\leq$2.5 \kms.
However, we found systematic shifts up to 1 pixel, corresponding to $\approx$30
\kms\ in radial velocity, for the two adopted grisms. To overcome the problem,
we adopted the approach suggested by \citet{kel03}. A model of the sky for each
slit was created and subtracted from the reduced 2D spectrum. This model was
also used to estimate and correct the systematic shifts from the daytime
wavelength calibration lamps, using night sky lines, mainly OI 5577.34 \AA\ ~for
the 1400V grism and a set of isolated skylines for the 1028z grism.

Yet another potential source of systematics in RV estimates was the centroiding
(see, e.g. \citealt{tolstoy01}). 
Thanks to the special care in the mask design, based on previous FORS2
preimaging of selected Carina fields, and to the very precise tracking of VLT,
this effect was limited at the level of 5-10 \kms.
To estimate the shift between the centroid of the star and the slit we adopted
the slit image which is acquired just before the spectrum. For each target the
centroid of the slit is evaluated by collapsing the slit along its width and
neglecting the pixels belonging to the star. The collapsed slit is then
subtracted away from the pixels belonging to the star, whose centroid is then
estimated with the task {\tt center} in IRAF. It is worth mentioning that we
also acquired slit images, for a couple of masks, soon after we collected the
spectra. These slit images were adopted to confirm that our estimates of the
centroids were minimally affected by tracking problems. 
The details of the pre-reduction and reduction of FORS2 spectra will be 
discussed in a forthcoming paper (Nonino et al.\ 2011, in preparation).

The GIRAFFE raw data were retrieved from the ESO Archive. These data were
reduced using IRAF. After the standard bias and flat correction, the spectra
were extracted using the traces from flats. Wavelength calibration is based on
daily calibration lamps: the formal solution rms was $<$0.02 \AA, corresponding
to a systematic $<$1 \kms\ in RV. After the fiber-spectra assignment, performed
using the table associated with the raw data, all the spectra were visually
inspected in order to remove bad spectra. A master sky for each exposure was
created from sky fibers, after the spectra had been cross-correlated to remove
shifts and scaled using the intensity of selected night sky lines in the
wavelength range 8250--8750 \AA, to better match the region with stellar
absorption lines. Subsequently the master skies were cross correlated with the
targets spectra, scaled in intensity and subtracted. The cross correlation among
the stacked skies showed that the relative accuracy in wavelength calibration
could give systematics of the order of 500 m s$^{-1}$ in RV. We finally coadded
all the individual spectra of the same target using the task {\tt scombine},
after correcting for the barycentric motion.

The GIRAFFE spectra taken with the high resolution grisms were reduced following a
similar approach.

\section{Radial velocity measurements}

To measure the RV of GMR and GHR spectra we fit individual spectral lines in
the coadded spectra. The RV of the LR spectra was measured on individual spectra
and the final RV was estimated as a weighted mean among the multiple spectra,
when available, of the same object.
All spectra were normalized to the continuum and the fit to single or multiple
lines performed using either a Gaussian or a Moffat function. The interactive
code developed to perform the RV measurements will be described in a forthcoming
paper (Fabrizio et al.\ 2011, in preparation). We selected by visual inspection
several spectral lines for each data set, namely H$_\beta$ and MgI triplet for
FORS2/1400V spectra (see the spectra plotted in the panels d) of Fig.~2), the
CaT for FORS2/1028z and GMR spectra (see the spectra plotted in the panels c)
and b) of Fig.~2) and several strong FeI lines for GHR spectra (see the spectra
plotted in the panels a) of Fig.~2). The quoted lines were typically fit with a
Moffat function with $\beta$=2, to properly account for the contribution of the
wings in the line fit. 

This preliminary estimate of the RV was validated by eye inspection and when
judged satisfactory it was adopted to perform an automatic estimate of the RV
using a large set of iron and heavy element lines. 
We adopted the line list for iron, $\alpha$- and heavy-elements recently
provided by \citet{roma08} and by \citet{pedi10}. The lines were selected 
according to the wavelength range and the spectral resolution of the different 
instruments and grisms. We ended up with a sub-sample of 30--70 lines for 
LR (grisms: 1028z, 1400V) spectra,  with $\approx$35 lines for GMR spectra 
and with $\approx$90 lines for the GHR spectra. For each spectrum we estimate 
the RVs as a weighted mean of the lines with the highest S/N ratio. This on 
average means 1-2 dozens of lines for LR and GMR spectra and approximately 
40 lines for the GHR spectra. 

The error in the RV based on individual lines was assumed equal to the sigma
of the fitting function\footnote{In particular, we assume
$e_{RV}$=c$\sigma_\lambda$/$\lambda$ \kms}. The RV of the entire spectrum was
estimated as a weighted mean over the different fitted lines. The error in the
radial velocity of coadded spectra (GMR,GHR) is the error of the weighted mean.
The RV of the LR spectra was estimated following the same approach for stars
with a single spectrum and as a weighted mean among the different mean
measurements for the stars with multiple spectra.

Data plotted in Fig.~3 show that the intrinsic error of the RV measurements
based on GHR spectra is on average (biweight mean)\footnote{Following the
referee's suggestion, we adopted the biweight location estimator \citep{andr72}, 
since it is a robust indicator insensitive to outliers in both Gaussian and 
non-Gaussian distributions. We adopted the definition given by \citep{beers90}, 
which includes data up to four standard deviations from the central location. 
This method is based on an iterative solution, but the process only requires a 
few steps to convergence.}  $<$1 \kms\ (panels a,b,c) and becomes of the order 
of 5 \kms\ for GMR03 spectra (panel d). Note that the intrinsic error of GMR08 
spectra is larger and $\sim$6 \kms. The difference is mainly due to the 
S/N ratio, since the latter sample is on average 0.5 mag fainter and its 
total exposure time is almost a factor of two shorter.

The intrinsic error of the RVs based on LR spectra is slightly larger and
ranges from $\sim$11 \kms\ (grism 1028z, panel f) to $\sim$10 \kms\ (grism
1400V, panel g). The difference between the FORS2 data sets is expected.
The first two lines of the CaT, at the typical RV of Carina stars, are
contaminated by atmospheric OH emission lines. Therefore, the RV measurements
based either on low-resolution spectra or on medium-resolution, low S/N ratio
spectra might be affected by slightly larger uncertainties in case the sky
lines are not precisely subtracted. The reader, interested in a more detailed
discussion, is referred to \citet{noni07} and to \citet{walker07}. Moreover, the
1028z spectra have a number of repeats that are approximately a factor of two
smaller than the number of repeats of the 1400V spectra (see \S 2). 
%

The fraction of objects for which we measured the RV is quite high, and
indeed the fraction ranges from $\sim$98\% for GHR and GMR03 spectra, to
$\approx$90\% for LR (1400V) spectra and to more than $\approx$80\% for
GMR08 and LR (1028z) spectra. Typically the stars with no RV measurements
have either low S/N or noisy spectra (see \S 2) or missing identification
in the photometric catalog. Eventually, the RV was measured in 2,165 out of
the 2,323 GIRAFFE spectra ($\sim$93\%) and in 381 out of the 429 ($\sim$89\%)
FORS2 spectra (see Fig.~3). In total the RV was measured in 1,812 stars using
GIRAFFE spectra and in 324 stars using FORS2 spectra (see Fig.~1). The final 
catalog includes RV measurements for 1,979 stars. 
%

To validate the approach adopted to measure the RVs, we decided to use
real spectra. This approach has several indisputable advantages when compared
with artificially generated spectra, since the different spectroscopic data
sets we are dealing with partially overlap. We can constrain the precision of
RVs based on HR spectra, since for each object we have three spectra with
similar spectral resolutions and covering three different wavelength regions
(HR10, HR13, HR14A, see \S 2). The panels a) and b) of Fig.~4 show the internal
comparison of our RV measurements as a function of the \vv-magnitude. The
difference and the RV dispersion of the entire sample and of the candidate
Carina stars (red dots, 180$\le$RV$\le$260 \kms, see \S 4) are minimal.

We can also constrain the precision of the GMR03 spectra, since this
subsample has more than three dozens of stars in common with the GHR spectra.
The panel c) shows the difference between the RVs based on the $<$GHR$>$ and
those based on the GMR03 spectra (weighted mean). The mean (biweight)
of the difference is 0.1 \kms\ for the candidate Carina stars (26) and smaller
than 1 \kms\ for the entire sample (35). The GMR03 RVs were corrected for the
difference.

The panel d) shows the difference between the RVs based on $<$GHR$>$ plus
GMR03 (weighted mean) spectra and RVs based on GMR08 spectra. We have more
than 120 stars in common and all of them are candidate Carina stars.
The difference is smaller than --3 \kms\ and the $\sigma$$\approx$8 \kms.
The difference with the RVs based on GMR03 spectra is mainly due to the fact
that the GMR08 spectra have a lower S/N ratio and are on average fainter. 
The GMR08 RVs were corrected for the difference. 
%

The panel a) of Fig.~5 shows the difference between the RVs based on the GIRAFFE
($<$GHR+GMR$>$, weighted mean) spectra and on the FORS2-1400V spectra (internal
comparison). The mean difference (biweight) over the entire sample of stars in
common (140/144) is --3.6 \kms\ and smaller (--2.7 \kms) for the candidate
Carina stars (122/125), while the dispersion is of the order of 12 \kms. The RVs
based on FORS2-1400V were accordingly corrected. The panel b) shows the
difference between the RVs based on GIRAFFE and on FORS2 ($<$GHR+GMR$>$ + 1400V,
weighted mean) and RVs based on FORS2-1028z spectra. The difference in RV and
dispersion are similar to the RVs based on the 1400V grism and they were also
corrected. Data plotted in panel c) show the difference between the RVs based on
the GIRAFFE ($<$GHR+GMR$>$, weighted mean) and on the FORS2 ($<$1028z+1400V$>$,
weighted mean) spectra after the corrections have been applied. The two samples
have more than 140 stars in common and the difference is $\sim$1 \kms\ for the
entire sample and vanishing for candidate Carina stars. 

The Fig.~6 shows the comparison between our RVs based on GIRAFFE spectra
--$<$GHR+GMR$>$-- with different RV measurements available in the literature
(external comparison).
The panel a) shows the comparison with the RVs based on echelle (+2D at LCO2.5m,
\citealt{mateo93}, crosses) and on multi-fiber spectra (HYDRA at CTIO 4m Blanco,
\citealt{majewski05}, filled circles).
The difference with the Mateo sample is quite small both in the mean (biweight)
and in the velocity dispersion. The difference with the Majewski sample is
larger ($\mu$=--3.6, $\sigma$=9.4 \kms), and it is caused by the lower S/N ratio
of the spectra they adopted to estimate the RVs. 

The panel b) shows the comparison between our RVs and RVs based either on 
high-resolution spectra collected with MIKE at Magellan (crosses) or on
medium-resolution spectra collected with GIRAFFE at VLT (filled circles)
provided by \citet{munoz06}.  The MIKE spectra were collected in slit mode
with a resolution of $\sim$19,000 (red arm) and the RV measurements are
based on the NIR calcium triplet.
The GIRAFFE spectra adopted by \citet{munoz06} are the same GMR03 spectra we
also included in this investigation. The difference with RVs based on MIKE  
spectra is minimal ($<$1 \kms), but the sample of stars in common is limited
(14). The difference with the RVs based on GIRAFFE spectra is larger $\sim$3
\kms\ if we account for the entire sample of stars in common (850/889 stars, 
$\sigma$=5.7 \kms) and for candidate Carina stars (353/373 stars, 
$\sigma$=6.7 \kms). The reasons for this difference are not clear, apart from 
the fact that we adopted a different approach to prereduce the spectra and to 
measure the RVs.

The panel c) shows the comparison with the RVs provided by W07. 
The RV measurements provided by W07 are based on high-resolution
(20,000--25,000) spectra collected with MMFS at Magellan. These spectra sample 
the region across the magnesium triplet (5140--5180\AA).
We have more than 550 candidate Carina stars in common and the data plotted in
this panel show that the two samples of RV measurements agree, within the errors
quite well. The difference is --2 \kms\ (biweight) and the RV dispersion is
smaller than 10 \kms. The same outcome applies to the RVs based on FORS2 spectra
(see panel d). We have more than 80 candidate Carina stars in common and the
difference is $\approx$1 \kms, while the dispersion $\sigma$$\sim$11 \kms.

We ended up with a data set including RV measurements for 1979
stars. For the objects with RVs based on both GIRAFFE and FORS2
spectra we performed a weighted mean, using as weights the
inverse square of the radial velocities standard errors of the
GMR, GHR and LR spectra. As a final test, we compared our entire
RV data set with the RVs provided by W07.
Data plotted in panel
e) further support the agreement between the two different sets of RV
measurements, and indeed the difference ranges from less than 1 \kms\ for the
entire sample (785 objects in common) to 2 \kms\ for candidate Carina stars
(574), while the dispersion is $\sim$9 \kms. The estimates of the mean
difference are based, once again, on the biweight and the number of neglected
objects is smaller than 9\%. The precision of both internal and external
validations, and in particular the good agreement with the RVs provided by W07,
support the approach we adopted to measure the RVs. 

The referee asked us to comment the impact of binaries stars on the current RV
measurements. The occurrence of binary stars among RG stars in dSph galaxies is
a highly debated topic and their impact on the RV dispersion ranges from a
sizable \citep{queloz95} to a small error \citep{olszewski96,hargreaves96} when
compared with the statistical error. However, in a recent investigation
\citet{minor10} found, by using a detailed statistical approach, that dSph
galaxies with RV dispersions ranging from 4 to 10 \kms can be inflated by no
more than 20\% due to the orbital motion of binary stars. It is worth noting
that the current LR spectra cover a time interval of three years (2004-2007) and
the number of repeats ranges from two to 14 (only eleven stars with a single
spectrum). The GMR spectra cover a time interval of five years (2003-2008) and
the number of repeats ranges from two to 35 (only 1 star with a single
spectrum). The GHR spectra were collected in 2005 and the number of repeats
ranges from seven to nine. The three different data sets have 145 objects in
common. This means that a significant fraction of stars in our sample have
spectra collected on a time interval of several years. Therefore, the current RV
measurements are less prone to significant changes caused by binary stars.
Moreover, the conclusions of the current investigation are minimally affected by
a possible uncertainty of the order of 20\% in the RV dispersion.

\section{Carina radial velocity distribution}

The top panel of Fig.~7 shows the RV distribution of the entire (GIRAFFE+FORS2)
data set. The well defined primary peak located at RV$\sim$220 \kms\ includes
candidate Carina stars. The current radial velocity distribution is soundly
supported by the radial velocity distribution recently provided by W07, but
based on a smaller number of stars (see the middle panel of Fig.~7). This
evidence and the minimal difference in the RV of the stars in common allowed us
to merge the two RV catalogs. 
%
For the objects in common weighted mean RVs were computed using as weights the
inverse square standard errors in the individual RV measurements. We ended up
with a sample of 2,629 stars (see the bottom panel of Fig.~7).

To further constrain the radial velocity distribution of the candidate Carina
stars we ran a Gaussian kernel on each star with a $\sigma$ equal to the RV
uncertainty. The solid red line plotted in the top panel of Fig.~8 was computed
by summing the individual Gaussians over the entire data set. We estimated the
mean and the $\sigma$ by fitting the smoothed RV distribution with a Gaussian
(dashed black line). We found that the peak (220.4$\pm$0.1 \kms) in the RV
distribution agrees quite well, within the errors, with similar estimates
available in the literature \citep{koch06,munoz06,walker07,walker09}. The same
outcome applies to the RV dispersion. Note that current estimate is based on a
homogeneous sample of candidate Carina stars (180$\leq$RV$\leq$260 \kms,
$\sim$4$\sigma$, 1208 stars) that is $\sim$55\% larger than any previous sample
of RVs. 

The middle panel of Fig.~8 shows the RV distribution of candidate Carina stars,
but based on the RV measurements (780 candidate Carina stars) provided by W07.
The two data sets provide, within the errors, very similar kinematic properties.
Note that by merging the two data sets of RV measurements we ended up, according
to the quoted selection criterion, with a sample of 1378 candidate Carina stars.
This sample of Carina RVs is $\approx$75\% larger than any previous RV sample.
To overcome subtle uncertainties in the estimate of the RV due to the possible
presence of outliers in the final data set, we estimated the biweight mean of
this sample and we found 220.9$\pm$0.1 \kms. This estimate is based on 1369
stars and this is the sample that we will adopt in the following to estimate the
RV distribution (red line in the bottom panel of Fig.~8). The mean based on the
Gaussian fit (dashed line) agrees quite well with the biweight mean. However,
the tails of the smoothed RV distribution appear shallower than the tails of the
fitting Gaussian. There is also evidence that the low-velocity tail might be
shallower than the high-velocity one. To provide a more quantitative estimate,
we estimated the $\chi^2$ of the two curves, and we found that the hypothesis
that the smoothed distribution is Gaussian can be discarded at 99\% confidence
level.  

The referee has expressed concerns about the possible dependence of the Carina
RV distribution on the different samples of RV measurements. To constrain this
effect we estimated the RV distribution of Carina stars using stars brighter
than \vv=20.25 mag (651). We fit the smoothed distribution with a Gaussian and
we found a mean RV $\mu$=221.3$\pm$0.1 \kms\ and a $\sigma$=8.7$\pm$0.1 \kms.
We also performed a new test including only stars brighter than \vv=20.50 mag
(1050) and we found $\mu$=220.9$\pm$0.1 \kms\ and a $\sigma$=8.9$\pm$0.1 \kms.
The quoted values agree quite well with the estimates based on the entire
sample. The two RV distributions based on smaller samples of brighter objects
also show asymmetric wings when compared with a Gaussian distribution.

To further constrain the plausibility of the selection criterion to pin point
candidate Carina stars, we plotted the kinematic candidates in the \vv,\bmi\
(left panel of Fig.~9) and in the \vv,\vmi\ (right panel of Fig.~9) CMDs. The
symbols of the different spectroscopic data sets are the same as in Fig.~1. Data
plotted in this figure show quite clearly the accuracy of the kinematic
selection, and indeed the bulk of the candidate Carina stars are located along
the expected evolutionary sequences (RG, RC, HB, blue plume). The new sample
also provides a clear separation between RG and Asymptotic Giant Branch (AGB,
\vv$\sim$20, \bmi$\sim$1.6--1.9 mag) stars.
Moreover and even more importantly, data plotted in Fig.~9 soundly support the
results recently provided by \citet{bono10} concerning the metallicity
distribution of Carina stars. The quoted authors found, using a color-color
plane (\umv, \bmi) to select candidate Carina stars, that the spread in color of
RG stars is significantly smaller than suggested by spectroscopic measurements.
The very narrow distribution in color of candidate Carina RGs is further
supported by the kinematic selection. There are a few RGs that attains colors
redder than typical Carina RGs. They might be either variables stars or
misidentified objects.

\section{Discussion}

In order to constrain whether the occurrence of asymmetries in the radial
velocity distribution is caused by the presence of substructures in Carina, we
investigated the change in the RV distribution as a function of the projected
radial distance ($\rho$). To avoid spurious fluctuations in the mean RV, we
ranked all the candidate Carina stars as a function of $\rho$ and estimated the
mean (biweight) and the median using the first 200 objects. The $\rho$ of each
subsample was estimated as the mean over the individual distances of the same
200 stars. We estimated the same quantities by moving of one object in the
ranked list until we accounted for the most distant 200 stars in our sample
located inside the tidal radius ($r_t$$\sim$28.8$\pm$3.6\arcmin,
\citealt{mateo98araa}). The error on the mean RV for individual bins is smaller
than one tenth of \kms. In order to provide robust constraints on the possible
uncertainties introduced by the number of stars per bin and by the number of
stepping stars we performed a series of MonteCarlo simulations. The estimated
mean dispersion is plotted as cyan shaded area across the RV mean in the bottom
panel of Fig.~10. Note that changes in the binning criteria can change the
extent of the secondary features, but their relative positions are minimally
affected.

A glance at the data plotted in the bottom panel of Fig.~10 shows that the mean
RV attains a well defined minimum --RV$\sim$219.5$\pm0.4$ \kms-- for
$\rho\sim$7\arcmin\ ($\sim$200 pc)\footnote{Projected distance on Carina were
estimated assuming a true distance modulus of 20.10$\pm$0.12 mag
\citep{dallora03,pietrynski09} together with the inclination angle and the axes
ratio \citep{mateo98araa}.} from the galaxy center and increases up to
$\sim$222.6$\pm$0.4 \kms\ for $\rho\sim$13\arcmin\ ($\sim$400 pc, 6$\sigma$
level) and a plateau of $\sim$221.4$\pm$0.4 \kms\ at larger distances
(600 pc, 4$\sigma$ level). This evidence is robust, since the mean
(biweight, black line) and the median (blue line) attain very similar values and
intrinsic errors. To further constrain the change of the kinematic status across
the body of the galaxy we also estimated the radial velocity dispersion
($\sigma_{RV}$, solid line middle panel of Fig.~10). Note that to avoid subtle
uncertainties caused by asymmetric radial velocity distributions we also
estimated for each bin the semi-interquartile range (SIQ, dashed-dotted line in
the middle panel of Fig.~10). Data plotted in the middle panel of Fig.~10 show
that both the $\sigma_{RV}$ and the SIQ show very similar radial trends, but the
former dispersion parameter attains values that are $\approx$50\% larger. The
difference indicates that the velocity dispersion should be cautiously treated
in dealing with asymmetric RV distributions \citep{strigari10}. We also found
that the dispersion is larger in the innermost regions, decreases in coincidence
of the minimum in the mean RV and after a mild increase attains an almost
constant value in the outermost galaxy regions. 
Note that the cyan shaded areas across the RV dispersion and the SIQ display
the intrinsic error on individual bin according to MonteCarlo simulations.
There is a hint that the dispersion could increase in the outskirts, thus
supporting the finding by W07. However, more data are required to constrain the
change in the projected radial velocity dispersion across the tidal radius. 

Interestingly enough, the kurtosis plotted in the top panel of Fig.~10 shows a
mirror trend compared with the projected radial velocity dispersion and with the
mean RV. This evidence is suggesting that in the Carina regions where the mean
RV is smaller, the dispersion is also smaller and the radial velocity
distribution more centrally peaked. The opposite trend characterizes the regions
in which the mean RV is larger. The outermost regions show flat trends in the
quoted parameters. Evidence of kinematic and photometric substructures have
already been found in several LG dwarfs \citep{monelli03,kleyna04,walker09}. In
particular, \citet{battaglia08} found evidence of a radial velocity gradient in
Sculptor, that they interpreted as an indication of internal rotation. 
The current data indicate that the RV distribution across the body of Carina
shows a radial gradient and possible evidence of rotation. However, the RV
distribution is far from being uniform, possibly suggestive of the occurrence of
well defined spatial substructures.

The referee has expressed concerns about the approach we adopted to estimate 
the weighted mean between the GIRAFFE and the FORS2 spectra. To clarify this
key point, Fig.~11 shows the same data of Fig.~10, but the parameters plotted
in the left panels are only based on the GIRAFFE spectra, while those plotted
in the right panels are based on the weighted mean between GIRAFFE and FORS2
spectra. The trends of the plotted parameters as a function of the radial
distance are the same. They are only slightly more noisy due to the decrease
in the sample size.

To further constrain the spatial extent of possible kinematic substructures we
investigated the same parameters, but they were estimated as a function of right
ascension ($\alpha$) and declination ($\delta$). Data plotted in panel c) of
Fig.~12 show the marginal of the radial velocity along the right ascension axis.
We found that the mean RV, when moving from E ($\sim$750 pc from the galaxy
center) to W ($\sim$--1060 pc from the center), decreases from 221.9$\pm$0.4 to
219.8$\pm$0.4 \kms. The difference is equal to $\sim$2.1 \kms\ (4$\sigma$ level)
and increases by almost a factor of two ($\sim$4.5 \kms, 9$\sigma$ level) if we
take into account the absolute maximum ($\sim$223.3$\pm$0.4 \kms, $\sim$320 pc
from the center) and the absolute minimum ($\sim$218.8$\pm$0.4 \kms, $\sim$-280 pc
from the center) of the mean RV. 

It is worth mentioning that the increase in the mean velocity both in the
external regions and outside the galactic center appears to be in the same
direction of the Carina proper motion (large blue arrow in panel d) of Fig.~12,
\citealt{piatek03,walker09}). However, the uncertainty affecting current
estimates of the Carina proper motion is quite large (see small blue arrows). 
Note that the increase in the mean RV cannot be in the direction of the Galactic 
center (heavy dashed line). Moreover and even more importantly, we found that in
a region of 200 pc across the galactic center the RV shows a secondary maximum
that is strongly correlated with the velocity dispersion (panel b) and
anticorrelated with the kurtosis (panel a). 

The change in the mean RV along the declination axis is similar to the change in
the right ascension axis. The mean RV decreases, when moving from N to S (panel
e), from 221.9$\pm$0.4 ($\sim$630 pc from the center) to 219.8$\pm$0.4 \kms\
($\sim$-650 pc from the center). The difference is at least at 4$\sigma$ level,
while the difference between maximum and minimum (222.3$\pm$0.4 \kms, $\sim$238 pc; 
219.1$\pm$0.4 \kms, $\sim$41 pc) is more than 3 \kms ($\sim$7$\sigma$ level). 
The absolute minimum is 
located close to the center in the IV quadrant and associated to the absolute minimum 
in RA. The radial trend outside this region shows the quoted secondary maximum and a
constant smooth trend at larger distances from the center. The dispersion (panel
f) and the kurtosis (panel g) are once again correlated/anticorrelated with the
behavior of the mean RV. It is noteworthy, that the quoted findings are
minimally affected by the FORS2, GIRAFFE and W07 data sets adopted to estimate
the RVs of candidate Carina stars.

To explain the occurrence of a kinematically cold population close to the center
of Sextans, \citet{kleyna04} suggested, using also photometric radial
distributions, that it could be the aftermath of a merging with a globular
cluster. To further constrain the nature of the above kinematic peculiarities we
investigated the Carina radial distributions, using the same approach adopted by
\citet{monelli03} and the recent photometric catalog by \cite{bono10}. The
isocontour levels (panel b) and the marginal (panel a) along the right ascension
axis plotted in Fig.~13 show that the radial distribution is flat-topped across
the Carina center. Note that the bright field star HD48652 (\vv=9.14 mag)
minimally affects this feature, since it is located at 2\Min77 from the Carina
center. On the other hand, the radial distribution is symmetric and with a well
defined peak along the declination axis. We drew three vertical lines across the
center and the two secondary maxima. Interestingly enough, the same lines
include the secondary maximum detected along the right ascension axis in the
mean RV. These empirical evidence indicates that the regions across the galaxy
center might form a substructure probably reminiscent of a transition between a
bulge-like and/or a disk-like substructure.

\section{Summary and conclusions}

We presented new radial velocity measurements of Carina dSph stars based on
low-resolution spectra (R$\sim$2500) collected with the FORS2 multi-object slit
spectrograph at the VLT. The key advantage of the current sample is that for the
first time we collected spectra of old (HB, \vv$\sim$20.75 mag), intermediate
(RC, \vv$\sim$20.5 mag) and young (BP, \vv$\sim$22 mag) stellar tracers. The
bulk of the spectra were collected using the 1400V grism covering the $H_\beta$
and the MgI triplet. For a fraction (20\%) of these targets we also collected
spectra using the 1028z grism centered on the NIR CaT. We secured 856 spectra of
359 stars with $\sim$20 hours of exposure time.

The above data were complemented by medium (R$\sim$6500) and high-resolution
(R$\sim$18000--23000) spectra collected with the GIRAFFE multi-object fiber
spectrograph at the VLT. The targets of this spectroscopic data set are RG and
RC stars (\vv$\le$20.6 mag). The medium resolution spectra are centered on the
NIR CaT, while the high-resolution ones cover a broad wavelength region
(5300$\lesssim$$\lambda$$\lesssim$6700 \AA). In total this data set includes
20,488 spectra of 1931 stars collected with $\sim$168 hours of exposure time.

The radial velocity distribution based on the above spectra agrees quite well
with similar data available in the literature. By assuming candidate Carina
stars the objects with RVs ranging from 180 to 260 \kms\ (4$\sigma$) we ended up
with a sample of $\sim$1,210 stars. By fitting these objects with a Gaussian we
found a mean RV --$<$RV$>$=220.4$\pm$0.1 \kms-- and a velocity dispersion
--$\sigma$=11.7$\pm$0.1 \kms-- that agree, within the errors, with similar
measurements provided by \citet{walker07}. 

The above kinematic selection was firmly supported by the position of the
candidate Carina stars along the expected evolutionary sequences (RG, AGB, RC,
HB, Blue Plume) of the CMD. The narrow distribution in color of the candidate
Carina RG stars confirms the recent findings by \citet{bono10}, based on
photometric selection criteria, that the spread in metallicity of Carina stars
is smaller than suggested by spectroscopic measurements \citep{koch06}.

The current catalog of Carina RV measurements was complemented by similar
measurements recently provided by \citet{walker07}. The stars in common were
averaged and we ended up with a sample of $\sim$1,370 RV measurements of
candidate Carina stars that is $\approx$75\% larger than any previous Carina RV
sample. We found that the hypothesis that the Carina RV distribution is Gaussian
can be discarded at the 99\% confidence level. A more detailed investigation of
the RV across the body of the galaxy indicates that the mean RV changes from
$\sim$220 \kms\ at a distance of 7\arcmin\ ($\sim$200 pc) from the center to
$\sim$223 \kms\ at 13\arcmin\ (6$\sigma$ level) and attains an almost constant
value of $\sim$221--222 \kms\ at larger distances (3--5$\sigma$ level).
Moreover, we found that in the Carina regions where the mean RV is smaller the
dispersion also attains smaller values and the RV distribution is more centrally
peaked (larger kurtosis). 

The mean RV at large distances from the galactic center decreases, when moving
from E to W, by more than 2 \kms\ (4$\sigma$ level). The difference increases by
almost a factor of two ($\sim$4 \kms, 9$\sigma$ level) if we account for the
difference between the absolute maximum and the absolute minimum of the mean RV.
The change in the mean RV, when moving from N to S, is similar. The difference
increases from more than 2 \kms\ (4$\sigma$ level) in the outermost regions to
more than 3 \kms\ (7$\sigma$ level) across the galaxy center. The RV gradient
appears to be in the direction of the Carina proper motion. However, this
parameter is still affected by large uncertainties to reach a firm conclusion.

Furthermore, there is also evidence of a secondary maximum in RV across the
Carina center ($|D|$$\le$200 pc). The occurrence of a secondary feature across
the Carina center is also supported by the flat-topped radial distribution
(photometric catalog) along the right ascension axis. In particular, the two
secondary maxima cover the same galaxy regions of the secondary maximum in the
RV radial variation. These results, once independently confirmed, are probably
reminiscent of a substructure with transition properties.

Evidence of substructures in dwarf galaxies dates back to \citet{jerjen00} who
detected a spiral structure in the Virgo dE IC3328 using deep, ground-based
R-band images. This discovery was soundly confirmed by \citet{lisker07} who
found that a significant fraction of bright ($M_{\it B}<16$) early-type dwarfs
in Virgo are characterized by disk-like features (spiral arms, bars).
Furthermore, they also found that the properties of dE in clusters are also
correlated with environmental density. This evidence supports numerical
simulations of galaxy harassment indicating that late-type galaxies undergo a
significant transformation when accreted in a cluster \citep{mastropietro05,
kormendy09}. It was also suggested that dEs with disk-like features might be the
low-luminosity tail of normal disk galaxies \citep{lisker06}. This working
hypothesis was recently supported by new and accurate kinematical data of IC3328
suggesting that the observed velocity dispersion is the aftermath of two
distinct substructures, namely a thin stellar disk and a dynamically hot
component. However, no robust conclusion could be reached due to limited radial
coverage \citep{lisker09}.

Irregular kinematic properties have also been found in the recently discovered
ultra-faint dwarf galaxy --Willman~I-- by \citet{willman10} using both
photometric and spectroscopic data. They found that galaxy stars located in
innermost regions show a radial velocity offset of 8 \kms\ when compared with
the outermost ones. Moreover, they also found initial hints of asymmetries in
the radial velocity distribution, but their spectroscopic sample is too small to
reach firm conclusions.

If the complex kinematic properties of Carina will be supported by future more
accurate and deep data sets then the occurrence of disk-like features might be
considered a wide spread property of gas-poor dwarf galaxies. A detailed
comparison between observed and predicted radial velocity distribution and
radial velocity dispersion profile is required to constrain the nature of Carina
kinematics.

\acknowledgments
It is a real pleasure to thank an anonymous referee for his/her positive
comments on the results of this investigation and for his/her pertinent
suggestions and criticisms that helped us to improve the content and the
readability of the paper.
We would like to thank M. Lombardi, for useful discussions concerning kinematic
properties of dwarf galaxies and S. Moehler for several enlightening suggestions
concerning the radial velocity measurements of hot stars. We also acknowledge
the ESO Cerro Paranal staff for collecting the spectroscopic data in service
mode and the ESO User Support Department for useful suggestions in handling the
raw spectra.
One of us (GB) thanks IAC for support as a science visitor. This publication
makes use of data products from VizieR \citep{ochse00} and from the Two Micron
All Sky Survey, which is a joint project of the University of Massachusetts and
the Infrared Processing and Analysis Center/California Institute of Technology,
funded by the National Aeronautics and Space Administration and the National
Science Foundation. We also thank the ESO/ST-ECF Science Archive Facility for
its prompt support.


\newpage
\begin{figure}[!ht]
\begin{center}
\label{fig1}
\includegraphics[height=0.65\textheight,width=0.65\textwidth]{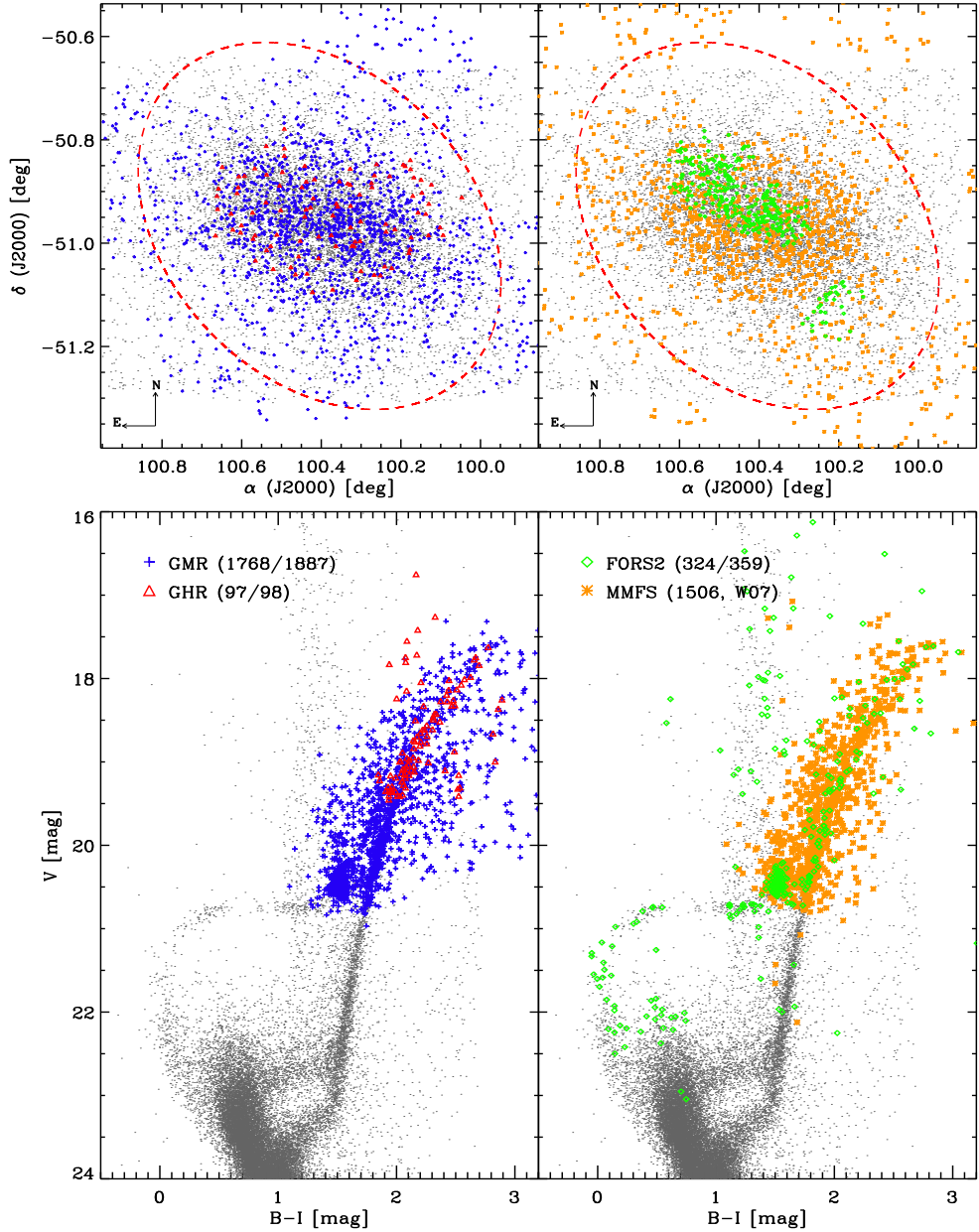}
\vspace*{0.75truecm}
\caption{Top Left: Spatial distribution of different spectroscopic data sets.
The blue pluses and the red triangles mark targets with medium (LR03,LR08) and
high-resolution (HR10,HR13,HR14A) GIRAFFE spectra. The background (grey dots)
is a composite Carina reference image based on randomly selected subsamples
of stars from the \citet{bono10} photometric catalog. The red dashed ellipse
shows the inclination ($i$=65\deg) and the radial extent (tidal radius,
$r_t$=28.8$\pm$3.6\arcmin\, \citealt{mateo98araa}).
Top Right: Same as the top left, but for targets with low-resolution FORS2
spectra (green diamonds). The yellow asterisks mark the positions of the targets
with high-resolution spectra collected with MMFS at Magellan by
\citet{walker07}.
Bottom Left: \vv,\bmi\ CMD of spectroscopic targets. The colored symbols are the
same as in the top left panel and show the position in the CMD of the targets
with GIRAFFE spectra. The numbers in parentheses are referred to the number of
stars with measured radial velocities and to the total number of stars with at
least one observed spectrum.
Bottom Right: Same as the bottom left, but for targets with FORS2 and MMFS
spectra.
}
\end{center}
\end{figure}

\begin{figure}[!ht]
\begin{center}
\label{fig2}
\includegraphics[height=0.65\textheight,width=0.65\textwidth, angle=90]{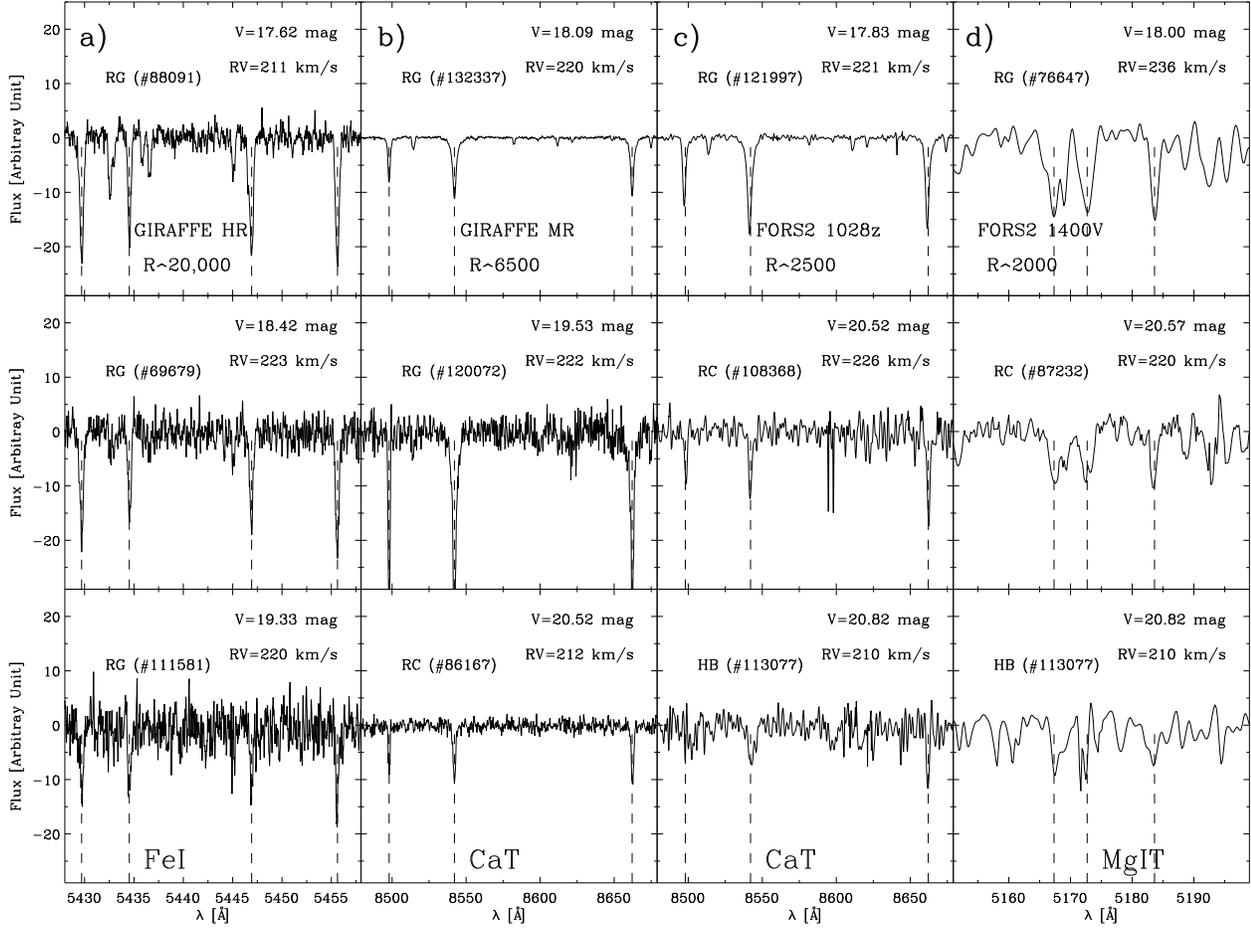}
\vspace*{0.75truecm}
\caption{Selected spectra collected with different spectrographs.
a): High-resolution spectra of three RGs, with apparent \vv\ magnitudes ranging
from $\sim$17.6 to $\sim$19.3, collected with GIRAFFE at VLT. The data plotted
in the three panels show a limited wavelength range (5428--5461 \AA) around four
strong FeI lines (vertical dashed lines). The individual radial velocities (RV)
are also labeled. The numbers in parentheses are referred to the IDs of
individual stars in the photometric catalog.
b): Same as panel a), but for medium-resolution spectra collected with GIRAFFE
at VLT. These spectra are centered on the NIR calcium triplet (CaT). The top and
the middle panel show the spectrum of two RGs, while the bottom panel the
spectrum of a RC star (\vv$\sim$20.5 mag).
%
c): Same as panel a), but for low-resolution spectra collected with FORS2
at VLT. These spectra are centered on the calcium triplet (CaT, grism: 1028z).
The top panel shows the spectrum of a RG star (\vv$\sim$17.8 mag), the middle
panel the spectrum of a RC star (\vv$\sim$20.5 mag) and the bottom panel the
spectrum of a HB star (\vv$\sim$20.8 mag).
d): Same as the panel c), but for low-resolution spectra centered on the
magnesium triplet (MgIT, grism: 1400V). The top panel shows the spectrum of a RG
star (\vv$\sim$18.0 mag), the middle panel the spectrum of a RC star
(\vv$\sim$20.6 mag) and the bottom panel the spectrum of a HB star
(\vv$\sim$20.8 mag).
}
\end{center}
\end{figure}

\begin{figure}[!ht]
\begin{center}
\label{fig3}
\includegraphics[height=0.65\textheight,width=0.65\textwidth]{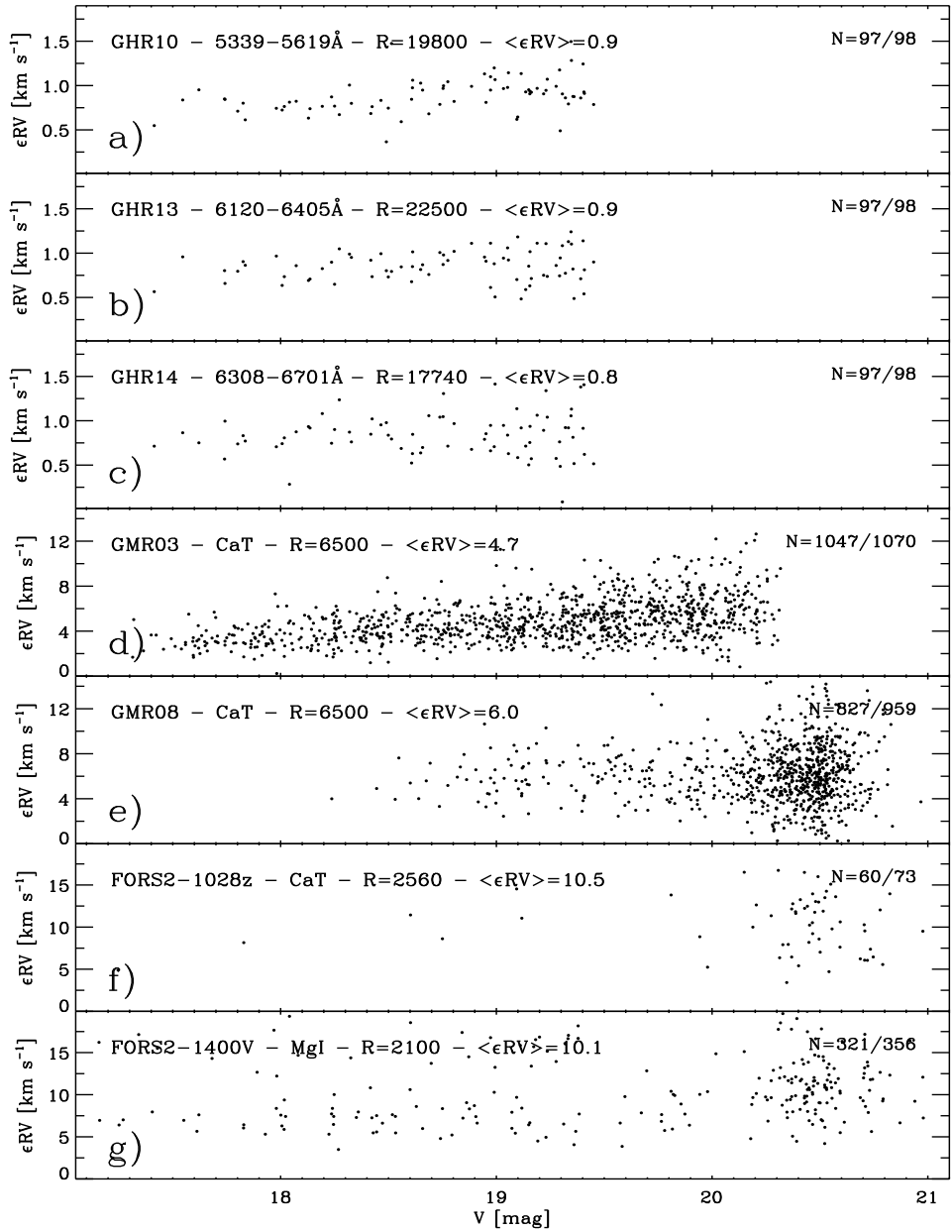}
\vspace*{0.75truecm}
\caption{Intrinsic errors in RV measurements of different spectroscopic data
sets as a function of visual magnitude. From top to bottom RV errors based on
GHR (panels a,b,c), on GMR (panels d,e) and on LR (panels f,g) spectra. In each
panel the labels give the wavelength range, the resolution of the spectra and
the biweight mean of the intrinsic error. The number of RV measurements and the
total number of spectroscopic targets are also labeled in the top right corner.
}
\end{center}
\end{figure}

\begin{figure}[!ht]
\begin{center}
\label{fig4}
\includegraphics[height=0.65\textheight,width=0.65\textwidth]{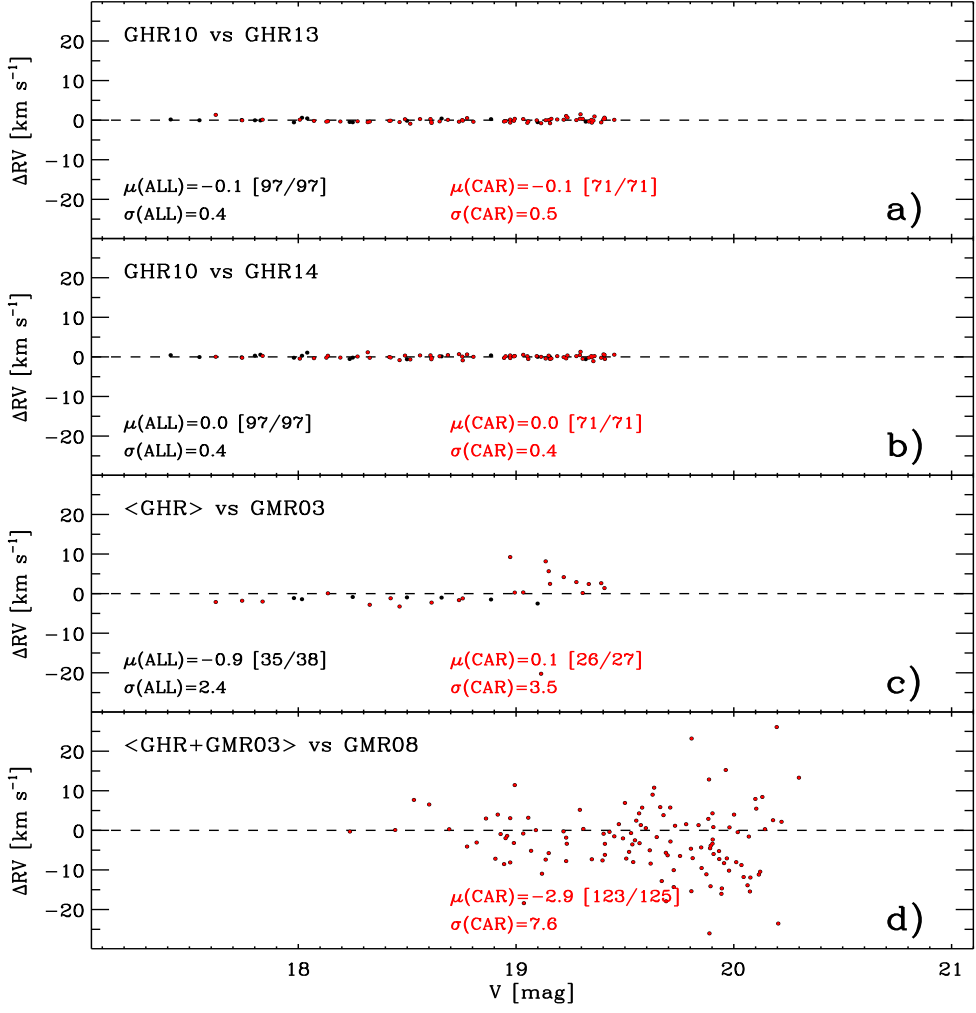}
\vspace*{0.75truecm}
\caption{Comparison of RV measurements based on medium- (GMR) and
high-resolution (GHR) spectra collected with GIRAFFE at VLT as a function of the
visual magnitude. The panels a) and b) show the difference in RV
($\Delta$RV=RV$_{10}$-RV$_{13,14}$) among the GHR spectra. The biweight mean
($\mu$), the standard deviation for the entire sample (black labels) and for the
candidate Carina stars (180$\le$RV$\le$260 \kms, red labels) are also labeled.
The numbers in square parentheses show the number of objects in common between
the two samples before and after the biweight mean.  
The panels c) and d): Same as top panels, but the difference is between the
weighted mean of the radial velocity based on the entire sample of GHR spectra
and on the GMR spectra.  
}
\end{center}
\end{figure}

\begin{figure}[!ht]
\begin{center}
\label{fig5}
\includegraphics[height=0.65\textheight,width=0.65\textwidth]{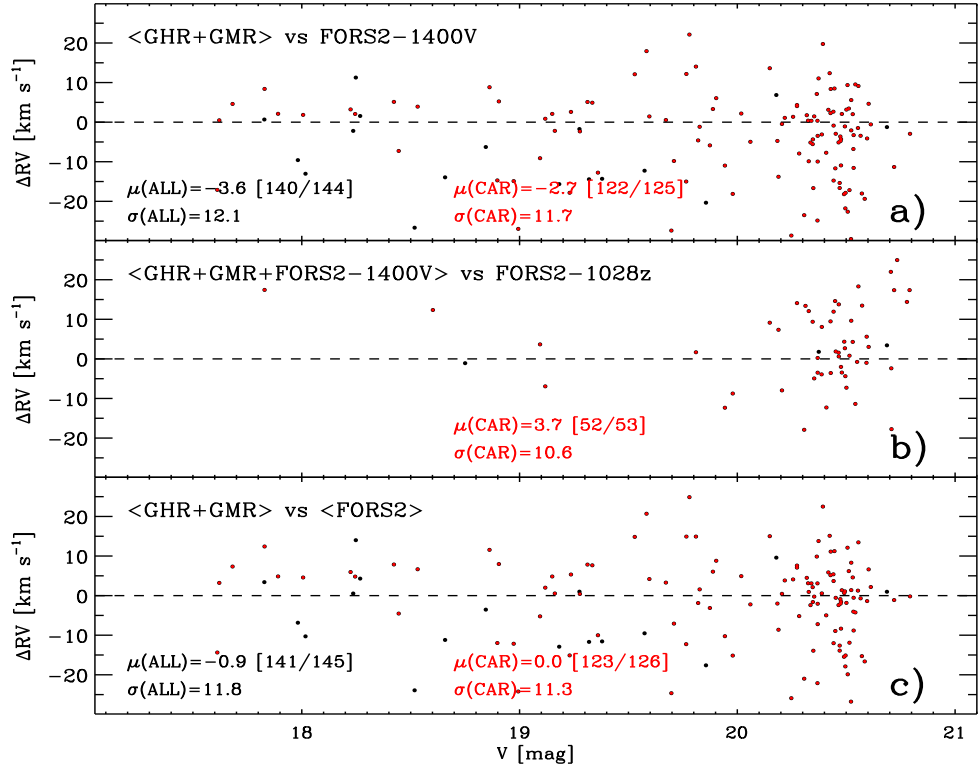}
\caption{Same as Fig.~4, but the difference is between the weighted mean of the
RV measurements based on all the GIRAFFE (GHR+GMR) spectra and those based on
the FORS2-1400V low-resolution (LR) spectra
($\Delta$RV=RV$_{GHR+GMR}$-RV$_{FORS-1400V}$, panel a). 
The panel b) Same as panel a), but the difference is between RVs based on
GIRAFFE plus FORS2-1400V spectra (weighted mean) and those based on the
FORS2-1028z spectra. 
The panel c) -- Same as panel a), but the difference is between the entire
sample of GIRAFFE and FORS2 RV measurements. The labels and the numbers in
parentheses have the same meaning of Fig.~4.
}
\end{center}
\end{figure}

\begin{figure}[!ht]
\begin{center}
\label{fig6}
\includegraphics[height=0.65\textheight,width=0.65\textwidth]{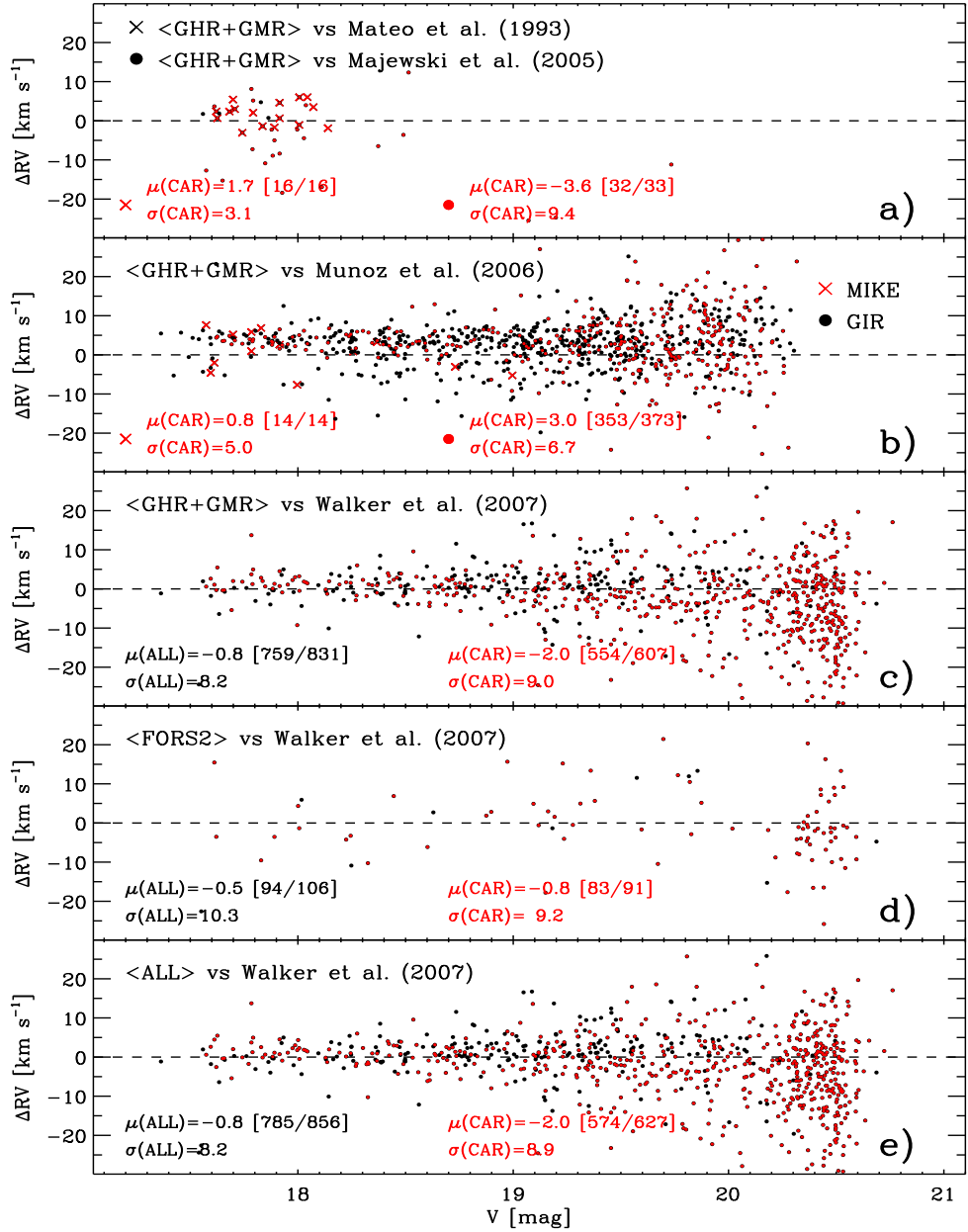}
\vspace*{0.75truecm}
\caption{Comparison among RV measurements based on different spectroscopic data
sets.The panel a) shows the difference between our GIRAFFE (GHR+GMR) RVs and
those provided by \citet{mateo93} and by \citet{majewski05}. The panel b) shows
the difference between our GIRAFFE (GHR+GMR) RVs and those provided by
\citet{munoz06} using two different data sets. The panel c) and d) show the
difference between our GIRAFFE (GHR+GMR) and FORS2 RVs with those provided by
W07 \citep{walker07} using spectra collected with MMFS at Magellan. The panel e)
shows the difference between our entire sample (GIRAFFE+FORS2) of RVs and those
provided by W07.
The biweight mean ($\mu$), the standard deviation for the entire sample (black
labels) and for the candidate Carina stars (180$\le$RV$\le$260 \kms, red labels)
are also labeled. The numbers in square parentheses show the number of objects
in common between the two samples before and after the biweight mean.  
} 
\end{center}
\end{figure}
\begin{figure}[!ht]
\begin{center}
\label{fig7}
\includegraphics[height=0.65\textheight,width=0.65\textwidth]{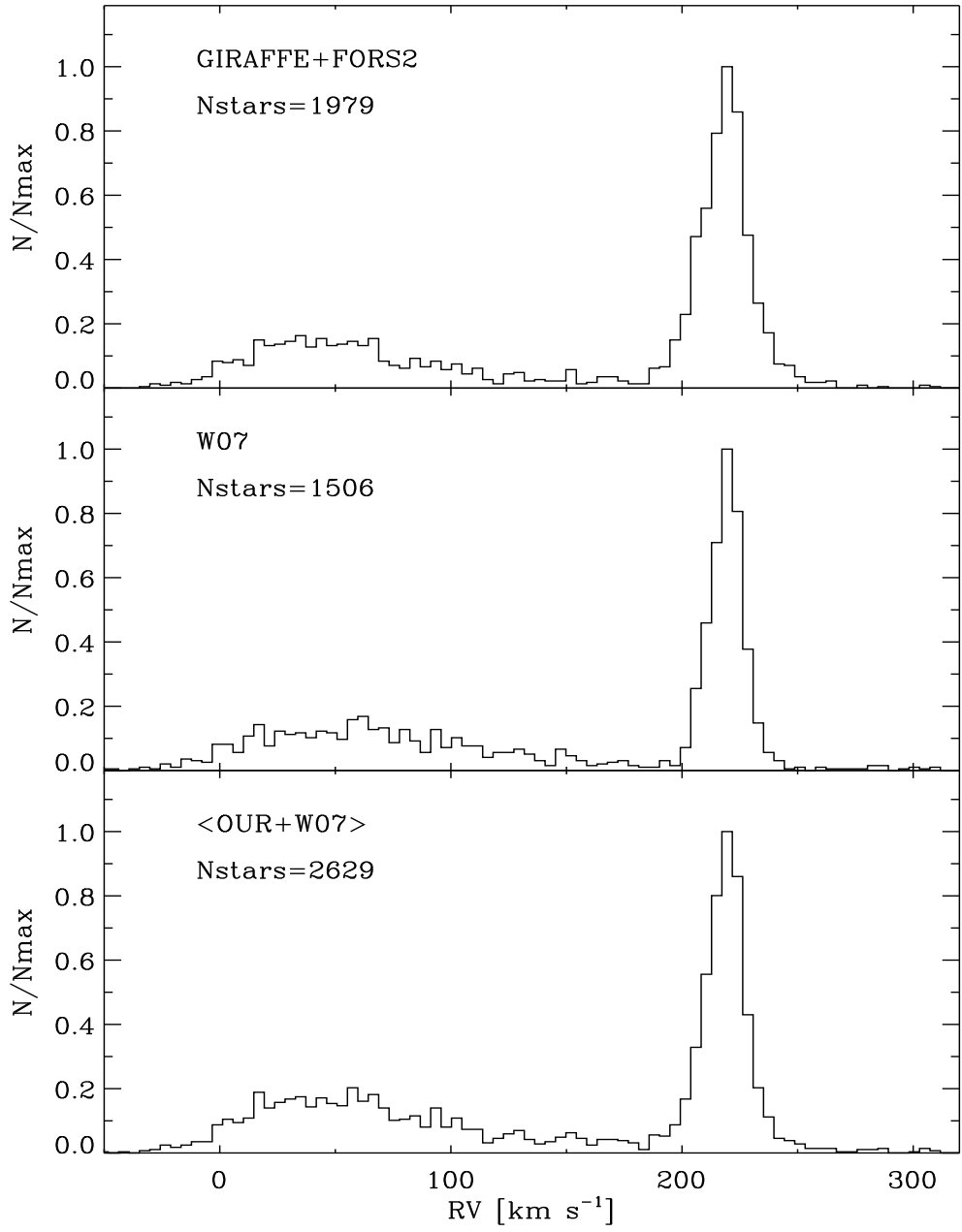}
\vspace*{0.99truecm}
\caption{Top: Radial velocity distribution of the entire sample normalized to
the maximum as a function of the radial velocity. Middle: Same as the top, but
based on RV measurements provided by W07 \citep{walker07}. Bottom: Same as the
top, but based on both our and W07 RV measurements. Note that for the stars in 
common in the two data sets we computed a weighted mean. See text for details.
}
\end{center}
\end{figure}
\begin{figure}[!ht]
\begin{center}
\label{fig8}
\includegraphics[height=0.65\textheight,width=0.45\textwidth]{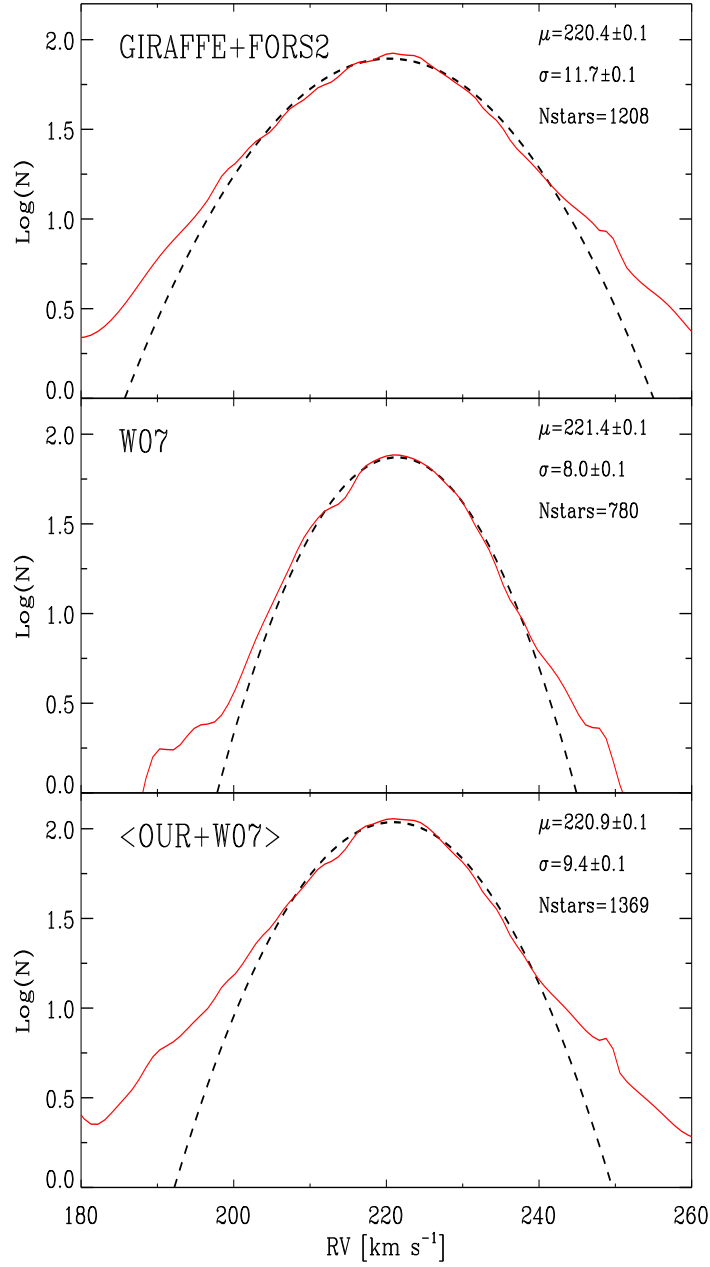}
\vspace*{0.99truecm}
\caption{Logarithmic radial velocity distribution for the candidate Carina stars
(180$\le$RV$\le$260 \kms, $\sim$4$\sigma$). The red solid line shows the
smoothed radial velocity distribution estimated running a Gaussian kernel on
individual RV measurements. The mean ($\mu$) and the $\sigma$ of the Gaussian
fit (dashed black line) are labeled together with the total number of candidate
Carina stars. Middle: Same as the top, but based on RV measurements provided by
W07 \citep{walker07}. Bottom: Same as the top, but based on both our and W07 RV
measurements.
}
\end{center}
\end{figure}
\begin{figure}[!ht]
\vspace*{1.5truecm}
\begin{center}
\label{fig9}
\includegraphics[height=0.5\textheight,width=0.65\textwidth,angle=90]{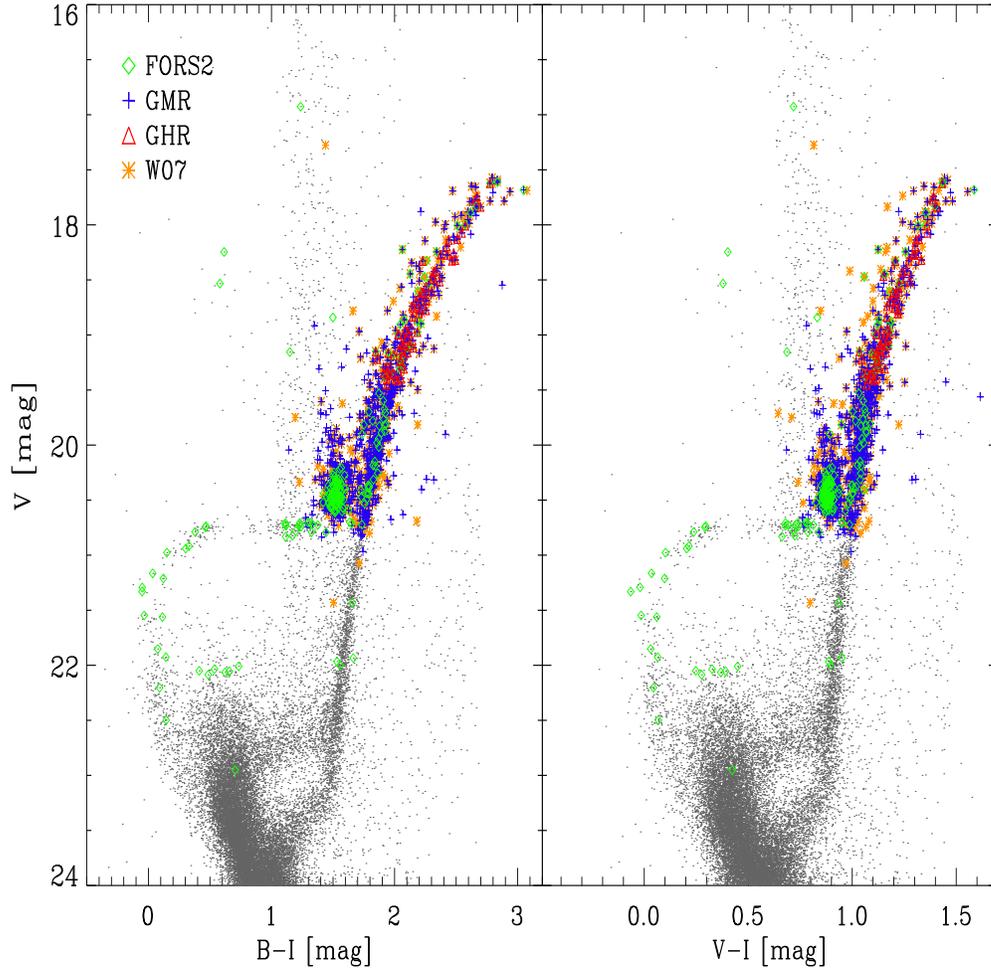}
\vspace*{0.99truecm}
\caption{\vv,\bmi (left) and \vv,\vmi (right) CMDs of candidate Carina stars
selected on the basis of the radial velocity (180$\le$RV$\le$260 \kms,
$\sim$4$\sigma$). The symbols of the different spectroscopic data sets are the
same as in Fig.~1.
}
\end{center}
\end{figure}

\begin{figure}[!ht]
\vspace*{1.5truecm}
\begin{center}
\label{fig10}
\includegraphics[height=0.5\textheight,width=0.65\textwidth]{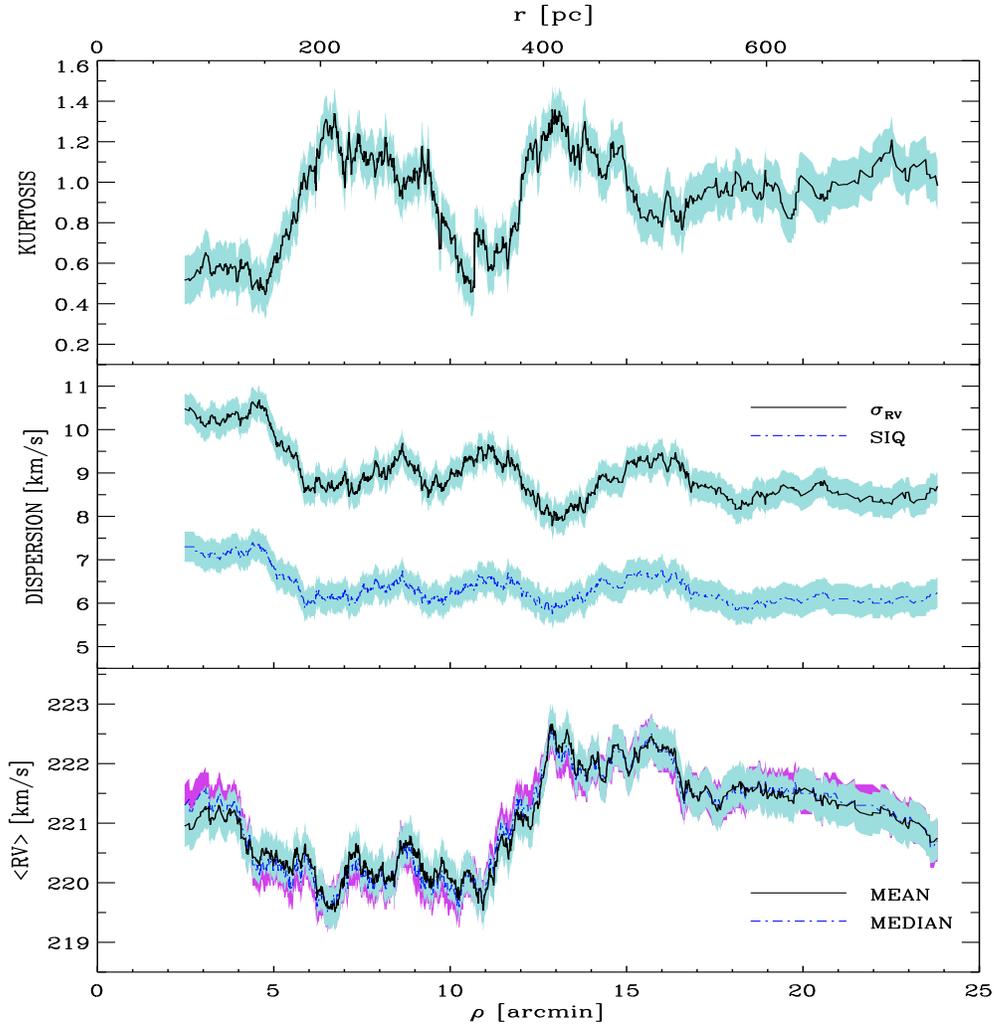}
\vspace*{0.99truecm}
\caption{Bottom: Mean (biweight, black line) and median (blue line) projected
mean radial velocity (\kms) as a function of radial distance in arcminutes
(bottom axis) and in parsecs (top axis). The individual values were estimated as
a running average over sub-samples of 200 stars (see text for more details). The
cyan and the purple shaded areas across the mean and the median curves display a
generous estimate of the intrinsic error on individual bins according to
MonteCarlo simulations. Middle: Same as the bottom, but for the projected radial
velocity dispersion ($\sigma_{RV}$, black line) and the semi-interquartile range
(SIQ, blue line). The cyan shaded areas across the curves display the intrinsic
error. Top: Same as the middle, but for the kurtosis.
}
\end{center}
\end{figure}

\begin{figure}[!ht]
\vspace*{1.5truecm}
\begin{center}
\label{fig11}
\includegraphics[height=0.5\textheight,width=0.65\textwidth]{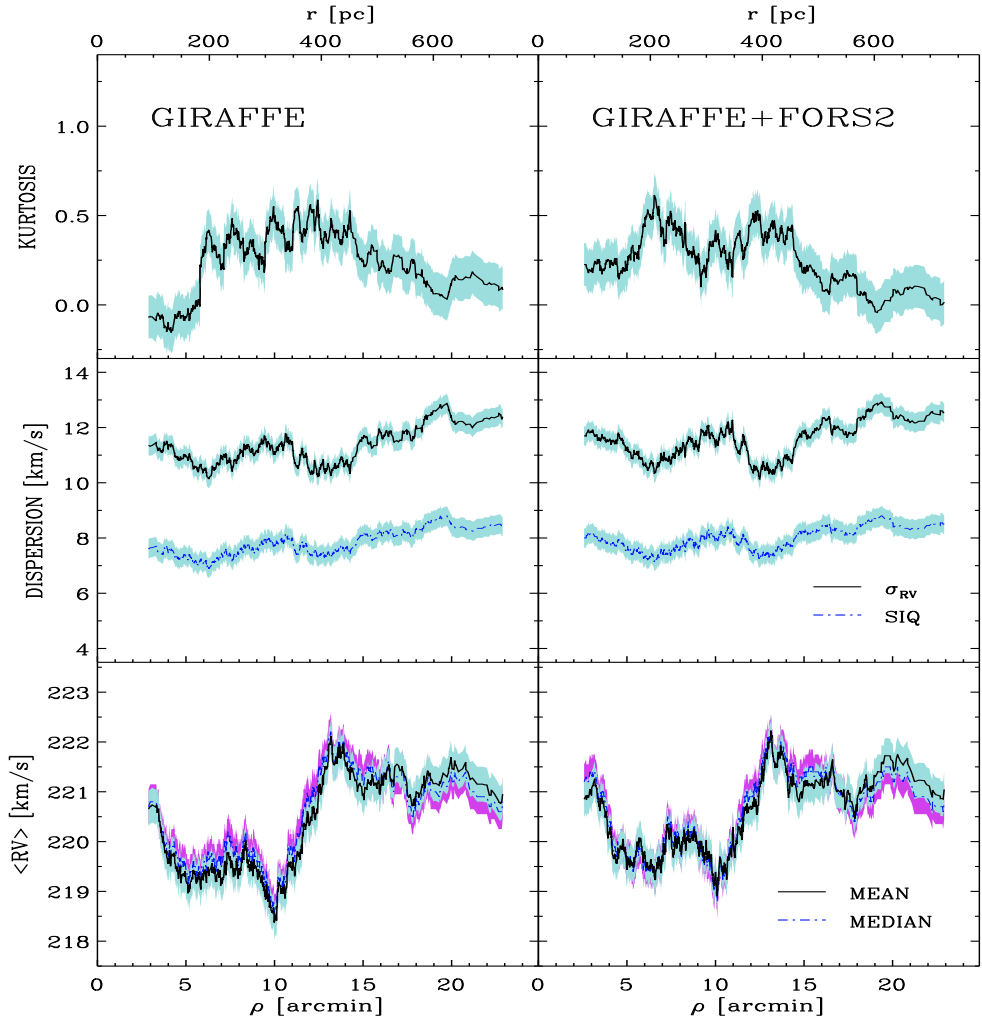}
\vspace*{0.99truecm}
\caption{
Left --- Same as Fig.~10, but the estimates of the different parameters are 
based only on  GIRAFFE ($GMR$+$GHR$) spectra.
Right --- Same as the left panels, but the estimates are based on the weighted 
mean between GIRAFFE and FORS2 spectra.
}
\end{center} 
\end{figure}

\begin{figure}[!ht]
\vspace*{1.25truecm}
\begin{center}
\label{fig12}
\includegraphics[height=0.65\textheight,width=0.75\textwidth]{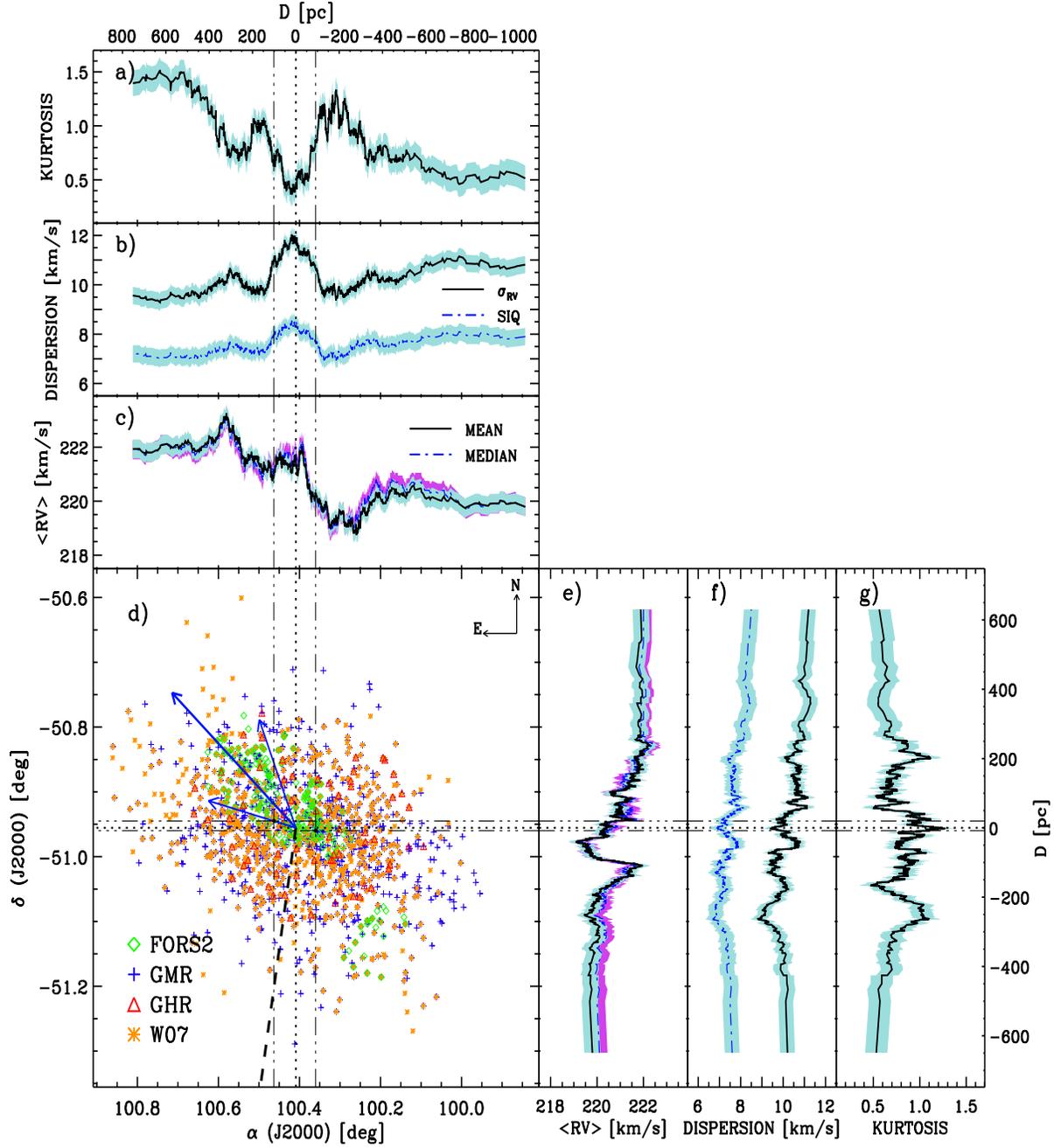}
\vspace*{0.88truecm}
\caption{
The panels a),b),c)-- Same as Fig.~10, but as a function of right ascension
($\alpha$, bottom axis) and distance (pc, top axis).
The panel d)-- Sky distribution of the different spectroscopic data sets.
Symbols and colors are the same as in Fig.~1. The long blue arrow shows the
Carina proper motion according to \citet{piatek03} and to \citet{walker09},
while the short ones display current uncertainties. The dashed black line shows
the direction of the Galactic center according to an observer located in the 
center of Carina. 
The dotted and the dashed-dotted thin lines display the secondary features 
identified in Fig.~10 and in Fig.~13. The panels e),f),g)-- Same as Fig.~10, 
but as a function of declination ($\delta$, left axis) and distance 
(pc, right axis).
}
\end{center}
\end{figure}

\begin{figure}[!ht]
\begin{center}
\label{fig13}
\includegraphics[height=0.55\textheight,width=0.70\textwidth]{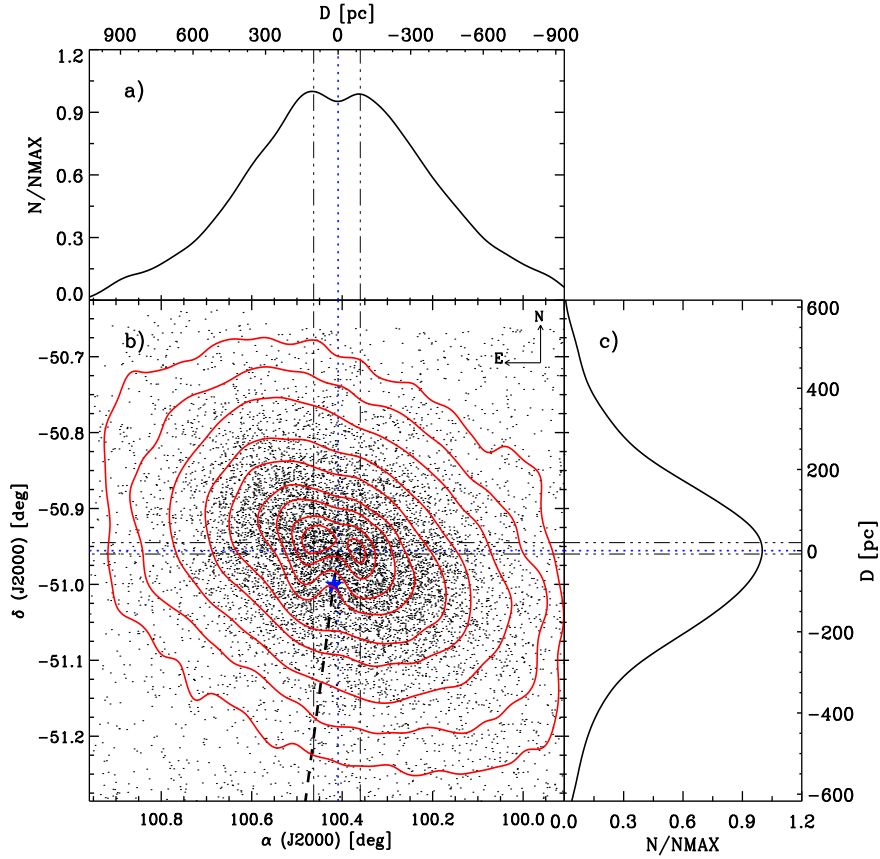}
\vspace*{0.88truecm}
\caption{
The panel a)-- Normalized radial profile along the right ascension axis of the
Carina photometric catalog \citep{bono10}.
The panel b)-- Sky distribution of the photometric catalog. Red contours display
iso-density levels ranging from 5 to 95\% with steps of $\sim$11\%. The two thin
dashed-dotted lines mark the two secondary peaks, while the thin dotted line
marks the center of the galaxy. The blue star shows the position of the bright
field star HD48652 (\vv=9.14 mag). The panel c)-- Same as the panel a), but
along the declination axis.
}
\end{center}
\end{figure}

\begin{deluxetable}{cccccrc}
\tabletypesize{\tiny}
\tablewidth{0pt}
\tablecaption{Log of FORS2 observations}
\tablehead{
\multicolumn{1}{c}{Date} &
\colhead{Pointing} &
\multicolumn{1}{c}{$\alpha$\tablenotemark{a}} &
\multicolumn{1}{c}{$\delta$\tablenotemark{b}} &
\colhead{Grism} &
\colhead{E.T.\tablenotemark{c}} &
\colhead{Seeing\tablenotemark{d}} \\
}
\startdata
 17-Feb-2004 & s3 & 06:41:41.0 & -50:56:03.6 & 1400V & 1360 & 0.69-0.79 \\
 17-Feb-2004 & l3 & 06:41:39.0 & -50:55:35.7 & 1400V & 4200 & 0.55-0.84 \\
 18-Feb-2004 & s3 & 06:41:41.1 & -50:56:03.3 & 1400V & 1360 & 0.75-0.80 \\
 18-Feb-2004 & l3 & 06:41:38.9 & -50:55:35.0 & 1400V & 4200 & 0.78-0.83 \\
 20-Feb-2004 & l4 & 06:41:27.3 & -50:57:13.0 & 1400V & 4200 & 0.43-0.76 \\
 20-Feb-2004 & s4 & 06:41:26.8 & -50:57:31.4 & 1400V & 1360 & 0.43-0.47 \\
 21-Feb-2004 & l4 & 06:41:27.4 & -50:57:12.6 & 1400V & 4200 & 0.52-0.68 \\
 21-Feb-2004 & l3 & 06:41:38.9 & -50:55:34.9 & 1400V & 4200 & 0.55-1.51 \\
 22-Feb-2004 & l5 & 06:40:51.0 & -51:07:56.4 & 1400V & 4200 & 0.77-0.83 \\
 23-Feb-2004 & s3 & 06:41:41.1 & -50:56:03.3 & 1400V & 1360 & 1.27-1.77 \\
 23-Feb-2004 & s2 & 06:42:00.5 & -50:50:27.3 & 1400V & 1360 & 1.26-1.59 \\
 13-Mar-2004 & s3 & 06:41:41.0 & -50:56:03.5 & 1400V & 1360 & 1.10-1.14 \\
 06-Dec-2004 & l1 & 06:42:04.0 & -50:53:21.2 & 1400V & 1800 & 0.35-0.66 \\
 06-Dec-2004 & l1 & 06:42:04.0 & -50:53:21.2 & 1028z & 1500 & 0.38-0.60 \\
 07-Dec-2004 & l1 & 06:42:04.0 & -50:53:21.2 & 1400V & 2700 & 0.49-0.73 \\
 07-Dec-2004 & l1 & 06:42:04.0 & -50:53:21.2 & 1028z & 900 & 0.51-0.54 \\
 08-Dec-2004 & s1 & 06:42:06.0 & -50:54:00.0 & 1400V & 1800 & 0.80-0.92 \\
 09-Dec-2004 & l2 & 06:42:04.7 & -50:49:57.3 & 1400V & 2400 & 0.42-0.44 \\
 09-Dec-2004 & l2 & 06:42:04.7 & -50:49:57.3 & 1028z & 1200 & 0.53-0.59 \\
 11-Dec-2004 & l2 & 06:42:04.7 & -50:49:57.3 & 1400V & 2400 & 0.55-0.56 \\
 14-Mar-2007 & f3 & 06:41:35.4 & -50:55:55.2 & 1400V & 8130 & 0.48-0.65 \\
 15-Mar-2007 & f3 & 06:41:35.5 & -50:55:55.1 & 1400V & 2710 & 0.70-1.67 \\
 15-Mar-2007 & f3r\tablenotemark{e} & 06:41:31.5 & -50:56:35.5 & 1400V & 5420 & 1.02-1.16 \\
 16-Mar-2007 & f3r\tablenotemark{e} & 06:41:31.5 & -50:56:35.6 & 1400V & 8130 & 0.51-0.95 \\
\enddata
\tablenotetext{a}{Right Ascension (J2000), the units are hours, minutes and seconds.}
\tablenotetext{b}{Declination (J2000), the units are degrees, arcminutes and arcseconds.}
\tablenotetext{c}{Exposure time (seconds).}
\tablenotetext{d}{Initial and final DIMM seeing (arcseconds) during the exposure time.}
\tablenotetext{e}{The pointings f3 and f3r are the same, but the latter was rotated by
180 degrees (Nonino et al.\ 2011, in preparation).}
\end{deluxetable}


\begin{thebibliography}{}
\bibitem[Appenzeller et al.(2008)]{appe98} Appenzeller et al.\ 1998, The Messenger 94, 1

\bibitem[Andrews et al.(1972)]{andr72} Andrees, D. F., Bickel, P. J., Hampel, F. R., Rogers, W. H. \& Tukey, J. W.\ 1972, Robust Estimates of Location: Survey and Advances (Princeton: Princeton University Press)

\bibitem[Battaglia et al.(2008)]{battaglia08} Battaglia, G., Helmi, A., Tolstoy, E., Irwin, M., Hill, V., \& Jablonka, P.\ 2008, \apjl, 681, L13

\bibitem[Beers et al.(1990)]{beers90} Beers, T.~C., Flynn, K., \& Gebhardt, K.\ 1990, \aj, 100, 32 

\bibitem[Belokurov et al.(2006)]{belokurov06} Belokurov, V., et al.\ 2006, \apjl, 647, L111

\bibitem[Bono et al.(2010)]{bono10} Bono, G., et al.\ 2010, \pasp, 122, 651 (paper III)

\bibitem[Bouchard et al.(2005)]{bouchard05} Bouchard, A., Jerjen, H., Da Costa, G.~S., \& Ott, J.\ 2005, \aj, 130, 2058

\bibitem[Cole(2009)]{cole09} Cole, A.~A.\ 2010, PASA, 27, 234 

\bibitem[Dall'Ora et al.(2003)]{dallora03} Dall'Ora, M., et al.\ 2003, \aj, 126, 197 (paper I)

\bibitem[Ferrarese et al.(2006)]{ferrarese06} Ferrarese, L., et al.\ 2006, \apjs, 164, 334

\bibitem[Gavazzi et al.(2005)]{gavazzi05} Gavazzi, G., Donati, A., Cucciati, O., Sabatini, S., Boselli, A., Davies, J., \& Zibetti, S.\ 2005, \aap, 430, 411

\bibitem[Hargreaves et al.(1996)]{hargreaves96} Hargreaves, J.~C., Gilmore, G., \& Annan, J.~D.\ 1996, \mnras, 279, 108 

\bibitem[Helmi et al.(2006)]{helmi06} Helmi, A., et al.\ 2006, \apjl, 651, L121

\bibitem[Jerjen \& Binggeli(1997)]{jerjen97} Jerjen, H., \& Binggeli, B.\ 1997, The Nature of Elliptical Galaxies; 2nd Stromlo Symposium, 116, 239

\bibitem[Jerjen et al.(2000)]{jerjen00} Jerjen, H., Kalnajs, A., \& Binggeli, B.\ 2000, \aap, 358, 845

\bibitem[Kelson(2003)]{kel03} Kelson, D. 2003 PASP, 115,688

\bibitem[Kirby et al.(2008)]{kirby08} Kirby, E.~N., Simon, J.~D., Geha, M., Guhathakurta, P., \& Frebel, A.\ 2008, \apjl, 685, L43

\bibitem[Kleyna et al.(2003)]{kleyna03} Kleyna, J.~T., Wilkinson, M.~I., Gilmore, G., \& Evans, N.~W.\ 2003, \apjl, 588, L21

\bibitem[Kleyna et al.(2004)]{kleyna04} Kleyna, J.~T., Wilkinson, M.~I., Evans, N.~W., \& Gilmore, G.\ 2004, \mnras, 354, L66

\bibitem[Koch et al.(2006)]{koch06} Koch, A., Grebel, E.~K., Wyse, R.~F.~G., Kleyna, J.~T., Wilkinson, M.~I., Harbeck, D.~R., Gilmore, G.~F., \& Evans, N.~W.\ 2006, \aj, 131, 895

\bibitem[Kormendy(1985)]{kormendy85} Kormendy, J.\ 1985, \apjl, 292, L9

\bibitem[Kormendy(1987)]{kormendy87} Kormendy, J.\ 1987, Structure and Dynamics of Elliptical Galaxies, 127, 17

\bibitem[Kormendy et al.(2009)]{kormendy09} Kormendy, J., Fisher, D.~B., Cornell, M.~E., \& Bender, R.\ 2009, \apjs, 182, 216

\bibitem[Lisker et al.(2006)]{lisker06} Lisker, T., Grebel, E.~K., \& Binggeli, B.\ 2006, \aj, 132, 497

\bibitem[Lisker et al.(2007)]{lisker07} Lisker, T., Grebel, E.~K., \& Binggeli, B.\ 2007, IAU Symposium, 235, 118

\bibitem[Lisker \& Fuchs(2009)]{lisker09} Lisker, T., \& Fuchs, B.\ 2009, \aap, 501, 429

\bibitem[{\L}okas et al.(2008)]{lokas08} {\L}okas, E.~L., Klimentowski, J., Kazantzidis, S., \& Mayer, L.\ 2008, \mnras, 390, 625

\bibitem[{\L}okas(2009)]{lokas09} {\L}okas, E.~L.\ 2009, \mnras, 394, L102

\bibitem[Majewski et al.(2005)]{majewski05} Majewski, S.~R.,Mu{\~n}oz, R.~R., Westfall, K.~B., \& Patterson, R.~J.\ 2005, Stellar Astrophysics with the World's Largest Telescopes, 752, 194

\bibitem[Mastropietro et al.(2005)]{mastropietro05} Mastropietro, C., Moore, B., Mayer, L., Debattista, V.~P., Piffaretti, R., \& Stadel, J.\ 2005, \mnras, 364, 607

\bibitem[Mateo(1998)]{mateo98araa} Mateo, M.~L.\ 1998, \araa, 36, 435

\bibitem[Mateo et al.(1993)]{mateo93} Mateo, M., Olszewski, E.~W., Pryor, C., Welch, D.~L., \& Fischer, P.\ 1993, \aj, 105, 510

\bibitem[Mateo et al.(1998)]{mateo98} Mateo, M., Hurley-Keller, D., \& Nemec, J.\ 1998, \aj, 115, 1856

\bibitem[Mateo et al.(2008)]{mateo08} Mateo, M., Olszewski, E.~W., \& Walker, M.~G.\ 2008, \apj, 675, 201

\bibitem[Minor et al.(2010)]{minor10} Minor, Q.~E., Martinez, G., Bullock, J., Kaplinghat, M., \& Trainor, R.\ 2010, \apj, 721, 1142 

\bibitem[Monelli et al.(2003)]{monelli03} Monelli, M., et al.\ 2003, \aj, 126, 218 (paper II)

\bibitem[Monelli et al.(2010a)]{monelli10a} Monelli, M., et al.\ 2010, \apj, 720, 1225

\bibitem[Monelli et al.(2010b)]{monelli10b} Monelli, M., et al.\ 2010, \apj, 722, 1864 

\bibitem[Mu{\~n}oz et al.(2006)]{munoz06} Mu{\~n}oz, R.~R., et al.\ 2006, \apj, 649, 201

\bibitem[Nonino et al.(2007)]{noni07} Nonino, M., et al.\ 2007, in The Future of Photometric, 
Spectrophotometric and Polarimetric Standardization, ed. C. Sterken, (San Francisco: ASP), 364, 295 

\bibitem[Ochsenbein et al.(2000)]{ochse00} Ochsenbein, F., Bauer, P., \& Marcout, J. 2000, A\&AS, 143, 221

\bibitem[Olszewski et al.(1996)]{olszewski96} Olszewski, E.~W., Pryor, C., \& Armandroff, T.~E.\ 1996, \aj, 111, 750 

\bibitem[Pasquini et al.(2002)]{pas02} Pasquini, L. et al.\ 2002, The Messenger 110, 1

\bibitem[Pedicelli et al.(2010)]{pedi10} Pedicelli, S., et al.\ 2010, \aap, 518, A11 

\bibitem[Piatek et al.(2003)]{piatek03} Piatek, S., Pryor, C., Olszewski, E.~W., Harris, H.~C., Mateo, M., Minniti, D., \& Tinney, C.~G.\ 2003, \aj, 126, 2346

\bibitem[Pietrzy{\'n}ski et al.(2009)]{pietrynski09} Pietrzy{\'n}ski, G., G{\'o}rski, M., Gieren, W., Ivanov, V.~D., Bresolin, F., \& Kudritzki, R.-P.\ 2009, \aj, 138, 459

\bibitem[Queloz et al.(1995)]{queloz95} Queloz, D., Dubath, P., \& Pasquini, L.\ 1995, \aap, 300, 31 

\bibitem[Romaniello et al.(2008)]{roma08} Romaniello, M., et al.\ 2008, \aap, 488, 731

\bibitem[Sanna et al.(2009)]{sanna09} Sanna, N., et al.\ 2009, \apjl, 699, L84

\bibitem[Skillman et al.(1989)]{skillman89} Skillman, E.~D., Terlevich, R., \& Melnick, J.\ 1989, \mnras, 240, 563

\bibitem[Strigari et al.(2010)]{strigari10} Strigari, L.~E., Frenk, C.~S., \& White, S.~D.~M.\ 2010, \mnras, 408, 2364 

\bibitem[Tolstoy et al.(2001)]{tolstoy01} Tolstoy, E. et al., 2001, MNRAS, 327, 918

\bibitem[Tolstoy et al.(2009)]{tolstoy09} Tolstoy, E., Hill, V., \& Tosi, M.\ 2009, \araa, 47, 371

\bibitem[Walker et al.(2007)]{walker07} Walker, M.~G., Mateo, M., Olszewski, E.~W., Gnedin, O.~Y., Wang, X., Sen, B., \& Woodroofe, M.\ 2007, \apjl, 667, L53 (W07)

\bibitem[Walker et al.(2009)]{walker09} Walker, M.~G., Mateo, M., \& Olszewski, E.~W.\ 2009, \aj, 137, 3100

\bibitem[Wilkinson et al.(2004)]{wilkinson04} Wilkinson, M.~I., Kleyna, J.~T., Evans, N.~W., Gilmore, G.~F., Irwin, M.~J., \& Grebel, E.~K.\ 2004, \apjl, 611, L21

\bibitem[Wilkinson et al.(2006)]{wilkinson06} Wilkinson, M.~I., Kleyna, J.~T., Wyn Evans, N., Gilmore, G.~F., Read, J.~I., Koch, A., Grebel, E.~K., \& Irwin, M.~J.\ 2006, EAS Publications Series, 20, 105

\bibitem[Willman et al.(2010)]{willman10} Willman, B., Geha, M., Strader, J., Strigari, L.~E., Simon, J.~D., Kirby, E., \& Warres, A.\ 2010, AJ, submitted, arXiv:1007.3499

\bibitem[Woo et al.(2008)]{woo08} Woo, J., Courteau, S., \& Dekel, A.\ 2008, \mnras, 390, 1453

\bibitem[Wyse(2010)]{wyse10} Wyse, R.~F.~G.\ 2010, AN, 331, 526

\bibitem[Zucker et al.(2006)]{zucker06} Zucker, D.~B., et al.\ 2006, \apjl, 650, L41

\end{thebibliography}
\end{document}